\documentclass[useAMS,usenatbib]{mn2e}

\usepackage{epsfig}
\usepackage{graphicx}
\usepackage{color}
\usepackage{array}
\usepackage{pstricks}
\usepackage{makeidx}
\usepackage{amsmath}
\usepackage{bm}
\usepackage{mathrsfs}
\usepackage{amssymb}
\usepackage{makeidx}
\usepackage{supertabular}
\usepackage{fancyhdr}
\usepackage{url}
\usepackage{natbib}
\usepackage{soul}

\usepackage{subfigure}
\usepackage{color, ulem}

\title[Forced inertial waves in rotating fluid bodies]{Nonlinear evolution of tidally forced inertial waves in rotating fluid bodies}
\author[B. Favier \textit{et al.}]{B. Favier$^{1,3}$\thanks{Email address: b.favier@damtp.cam.ac.uk}, A. J. Barker$^{2}$, C. Baruteau$^1$, and G. I. Ogilvie$^1$\\
$^{1}$Department of Applied Mathematics and Theoretical Physics, University of Cambridge, \\
\ Centre for Mathematical Sciences, Wilberforce Road, Cambridge CB3 0WA, UK \\
$^{2}$Center for Interdisciplinary Exploration and Research in Astrophysics (CIERA) \& Department of Physics and Astronomy, \\
\ Northwestern University, 2145 Sheridan Road, Evanston, IL 60208, USA \\
$^{3}$Centre for Mathematical Science, City University London, Northampton Square, London EC1V 0HB, UK}

\begin{document}

\date{\today}

\pagerange{\pageref{firstpage}--\pageref{lastpage}} \pubyear{2012}

\maketitle

\label{firstpage}

\begin{abstract}
We perform one of the first studies into the nonlinear evolution of tidally excited inertial waves in a uniformly rotating fluid body, exploring a simplified model of the fluid envelope of a planet (or the convective envelope of a solar-type star) subject to the gravitational tidal perturbations of an orbiting companion.
Our model contains a perfectly rigid spherical core, which is surrounded by an envelope of incompressible uniform density fluid.
The corresponding linear problem was studied in previous papers which this work extends into the nonlinear regime, at moderate Ekman numbers (the ratio of viscous to Coriolis accelerations).
By performing high-resolution numerical simulations, using a combination of pseudo-spectral and spectral element methods, we investigate the effects of nonlinearities, which lead to time-dependence of the flow and the corresponding dissipation rate.
Angular momentum is deposited non-uniformly, leading to the generation of significant differential rotation in the initially uniformly rotating fluid, \textit{i.e.}~the body does not evolve towards synchronism as a simple solid body rotator.
This differential rotation modifies the properties of tidally excited inertial waves, changes the dissipative properties of the flow, and eventually becomes unstable to a secondary shear instability provided that the Ekman number is sufficiently small.
Our main result is that the inclusion of nonlinearities eventually modifies the flow and the resulting dissipation from what linear calculations would predict, which has important implications for tidal dissipation in fluid bodies.
We finally discuss some limitations of our simplified model, and propose avenues for future research to better understand the tidal evolution of rotating planets and stars.
\end{abstract}


%
%
\section{Introduction}
Understanding the gravitational tidal interactions between two orbiting bodies is an important unsolved problem in astrophysics.
It is often the case that one or more of the bodies involved in the interaction is wholly or partly composed of fluid layers, such as a star or giant planet, an ice giant with a thick atmosphere, or a terrestrial planet with deep ocean.
The processes by which the orbits and spins of these bodies evolve due to tidal interaction are poorly understood.
The continuing discovery of short-period extrasolar planets makes a theory of tidal interactions very relevant, since they may have played an important role in producing the observed properties of these systems.

The importance of tidal interactions in a given system, quantified by the relevant timescale for tides to cause the orbit and spin of a body to evolve, is related to the rate at which tidal energy is dissipated.
The mechanisms responsible for this dissipation depend strongly on the internal structure of the body.
For fluid bodies, the hydrostatic response of a homogeneous fluid body to the varying gravitational potential of the companion is often referred as to the equilibrium tide \citep{darwin1880}.
In the convective envelope of a gas giant planet or low-mass star, the equilibrium tide is thought to be dissipated by the turbulent motions driven by thermal convection, though there are significant uncertainties in the efficiency of this process \citep{Zahn1966,GN1977,penev2007,penev2009,OL2012}.
In stably stratified fluid layers, such as the radiative layers of a planet or star, the resonant excitation of $g$-mode oscillations (or internal gravity waves) is thought to dominate the tidal dissipation \citep{zahn1970,GD1998}.

Since all astrophysical fluid bodies rotate, it is important to understand the effect that rotation has on the rates of tidal dissipation.
This is particularly important because rotating flows support oscillatory motions called inertial waves \citep{greenspan1968}.
These waves are restored by the Coriolis force and can be excited by low frequency tidal forcing when the absolute value of the forcing frequency (in a frame rotating with the fluid) is less than twice the spin frequency of the body.
The importance of these waves at contributing to tidal dissipation has been emphasised in recent years \citep{ogilvie2004,Wu2005,OgilvieLin2007,IvPap2007,GoodmanLackner2009,IvPap2010,papaloizou2010}.
In nearly adiabatically stratified convective regions, internal gravity waves are not supported and the excitation and dissipation of inertial waves might play a dominant role.

In the case of a full sphere of incompressible fluid, a complete set of smooth inviscid inertial modes exist, which can be derived analytically.
However, the case of a spherical shell is more relevant for the interiors of giant planets with solid cores, terrestrial planets with deep oceans, and, to some extent, the convective envelope of a solar-type star.
In this geometry, regular inviscid inertial modes do not exist, and the viscous eigenfunctions correspond to rays propagating along the characteristics of the Poincar\'e wave equation.
For some frequencies, these wave beams can converge toward limit cycles, known as wave attractors, as they alternately reflect from the inner solid core and the outer boundary \citep{rieutord1997,rieutord2001}.
As a result, the viscous dissipation rate has a very sensitive dependence on the tidal forcing frequency \citep{ogilvie2004,ogilvie2009,rieutord2010}.
Moreover, this dissipation is strongly enhanced for particular frequencies, hence it may play an important role in contributing to tidal evolution in rotating fluid bodies.

Most of the previous studies of inertial waves in a spherical shell have considered the linear behaviour of these singular inertial wave solutions.
However, as the Ekman number (the ratio of viscous to Coriolis accelerations) is decreased towards the values that it is thought to take in planetary or stellar interiors, and as the amplitude of the tidal forcing is increased, nonlinearities will play an increasingly important role.
When a wave attractor is present, it is known from experimental and theoretical studies that the inertial wave beam can become unstable and break, leading to a turbulent flow and a different mechanism of dissipation \citep{Scolan2013,jouve2013}.
The dissipation rate may be modified somewhat from that predicted by linear theory.

A related nonlinear mechanism that could contribute to tidal dissipation is the so-called elliptical instability \citep{Pierrehumbert1986,bayly1986,waleffe1990,kerswell2002}.
This is an instability of elliptical streamlines, such as those in the equilibrium tidal flow in a rotating fluid body.
The elliptical instability leads to the excitation of inertial waves through a parametric resonance.
The nonlinear outcome of this instability in isolation has been studied in laboratory experiments (e.g. \citealt{lacaze2004,lebars2007,lebars2010}), as well as local \citep{barker2013a,barker2013b} and global numerical simulations (e.g. \citealt{cebron2010,cebron2013}).
These works suggest that the elliptical instability could contribute to tidal dissipation at short orbital periods.
However, the elliptical instability has yet to be studied in detail in spherical shell geometry (except briefly in \citealt{cebron2010}).

In addition, nonlinearities could generate zonal (net azimuthal) flows, which might also alter the dissipative properties of the flow.
An initial study into the effects of cylindrical differential rotation i.e. zonal flows, on inertial waves in a spherical shell has been undertaken by \cite{baruteau2013}.
They show that a differential rotation can modify the path of characteristics, and alter the frequency-dependence of the viscous dissipation.
In the case of precessing or librating flows, it is well known that zonal flows can be driven by nonlinearities in the Ekman boundary layers \citep{busse1968,noir2001,calkins2010,sauret2013}.
The weakly nonlinear model of \cite{tilgner2007} also illustrated a mechanism by which nonlinearities of inertial waves in a spherical shell can generate zonal flows even in the absence of Ekman boundary layers.
Intense axisymmetric flows were observed experimentally in the case of a rotating deformed sphere \citep{morize2010}.
Although the detailed mechanisms that generate them depend on the type of forcing and on the boundary conditions, the existence of zonal flows is a generic feature of forced rotating flows, hence it is important to study their effect on tidal dissipation in fluid bodies.

In this paper, we consider a simple model of a rotating fluid body subject to tidal forcing first studied in \cite{ogilvie2009}.
We explore the nonlinear behaviour of this model using high-resolution three-dimensional numerical simulations. 
To our knowledge, this is one of the first attempts to numerically compute forced inertial waves in a spherical shell as an initial value problem (see also \citealt{papaloizou2010}), and the first to study the nonlinear problem. Our model and the numerical methods adopted are presented in Section~\ref{sec:model}.
We first illustrate the model and compare with previous results in the linear regime in Section~\ref{sec:linear}.
The main results concerning the nonlinear regime are discussed in Section~\ref{sec:nonlinear}.
Finally, our conclusions and future directions are given in Section~\ref{sec:conclusion}.

%
%
\section{Physical model}\label{sec:model}

\subsection{Model and governing equations}

Our model is based on \cite{ogilvie2009}.
We suppose that the body is in uniform rotation with angular velocity $\bm{\Omega}=\Omega\bm{e}_z$ where $\bm{e}_z$ is the unit vector in the vertical direction.
The fluid is incompressible with uniform density $\rho$, henceforth taken to be unity, without loss of generality.
This assumption is adopted at this stage for simplicity.
We neglect centrifugal distortion, effectively limiting us to studying bodies that rotate much slower than their dynamical frequency.

We define a spherical polar coordinate system centred on the body with coordinates $(r, \theta, \phi)$ in the rotating frame.
We aim to study the response of the fluid contained within a spherical shell with $r_i<r<r_e$ to a tidal gravitational potential $\Psi \propto A \Re \left[ Y_{l}^{m}(\theta,\phi) e^{-\textrm{i}\omega t}\right]$.
$Y_l^m$ is the spherical harmonic of degree $l$ and order $m$ with $l \ge m$, $\omega$ is the tidal frequency in the rotating frame, related to that in the inertial frame $\omega_i$ by $\omega=\omega_i-m\Omega$.
$A$ is an arbitrary real amplitude.
Our system of units is defined such that $\rho=1$, $\Omega=1$ and $r_{e}=1$. 
The equations of motion are then
\begin{align}
\label{eq:momentum}
\frac{\partial\bm{u}}{\partial t}+\bm{u}\cdot\nabla\bm{u}+2\bm{e}_{z}\times\bm{u} & = -\nabla W + E\nabla^2\bm{u}\\
\label{eq:incomp}
\nabla\cdot\bm{u} & = 0 \ ,
\end{align}
where 
\begin{equation}
E=\frac{\nu}{\Omega r_{e}^{2}}
\end{equation} 
is the Ekman number (the ratio of viscous to Coriolis accelerations), $\nu$ is kinematic viscosity of the fluid, assumed to be uniform.
$W=p+\Psi+\Phi'$, where $p$ is the pressure, $\Psi$ is the external tidal potential and $\Phi'$ is the internal self-gravitational potential perturbation.
$\Phi'$ is an order-unity multiple of $\Psi$, it is in fact $(3/2)\Psi$ for a homogeneous sphere, so it just amplifies the tidal potential and increases the amplitude of the tide.

The linear properties of inertial waves depend strongly on the size of the core (e.g. the dissipation rate appears to scale as the fifth power of the core size -- \citealt{GoodmanLackner2009,ogilvie2009}).
\textcolor{black}{Throughout this paper, the core size is however fixed to $r_{i}=0.5$ in order to reduce the number of parameters.
This core size is likely to limit the appearance of the elliptical instability, which is not the main focus of this work.
}
This is because the nonlinear coupling between the equilibrium tide and the singular inertial waves in a spherical shell may be weaker than with the regular inertial modes of a core-free body.
In addition, we will only consider moderate amplitude tidal forcing in this paper, with $A \ll1$.
At these amplitudes, it is likely that the elliptical instability will be weak, since the maximum (inviscid) growth rate of the elliptical instability is proportional to the amplitude of the forcing ($\sim A/\Omega$).
Choosing such a small $A$ is therefore likely to eliminate the elliptical instability from playing a major role in our calculations.
It may also mean that it is difficult to observe any parametric subharmonic instabilities of the inertial wave beams, which are likely to limit the wave amplitudes when $E\rightarrow 0$.
We return to this point later in Section~\ref{sec:nonlinear}.

\subsection{Boundary conditions}

\begin{figure}
  \begin{center}
    \begin{tabular}{c}
      \hspace{8mm}\resizebox{80mm}{!}{\includegraphics{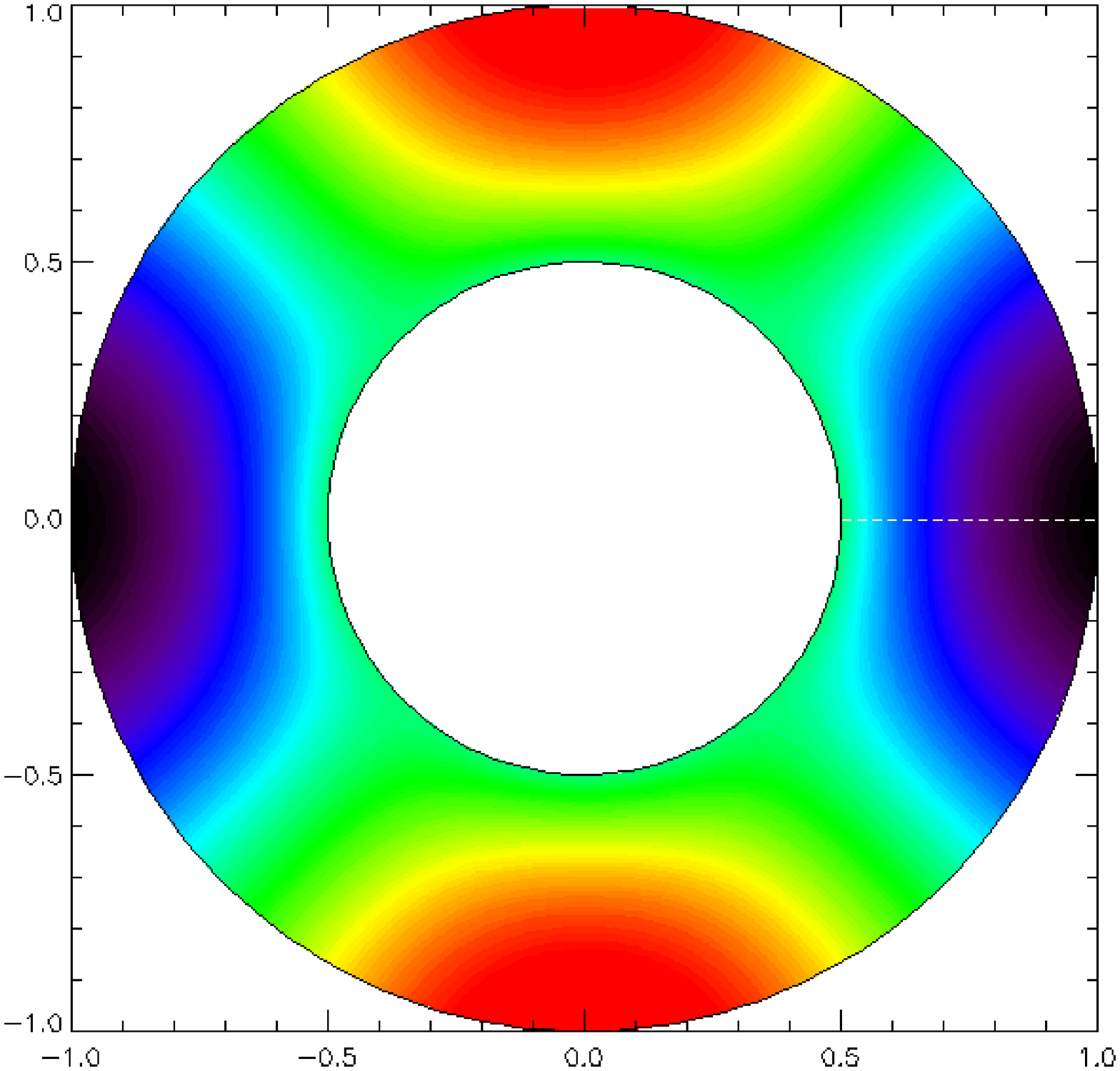}}\\
      \resizebox{80mm}{!}{\includegraphics{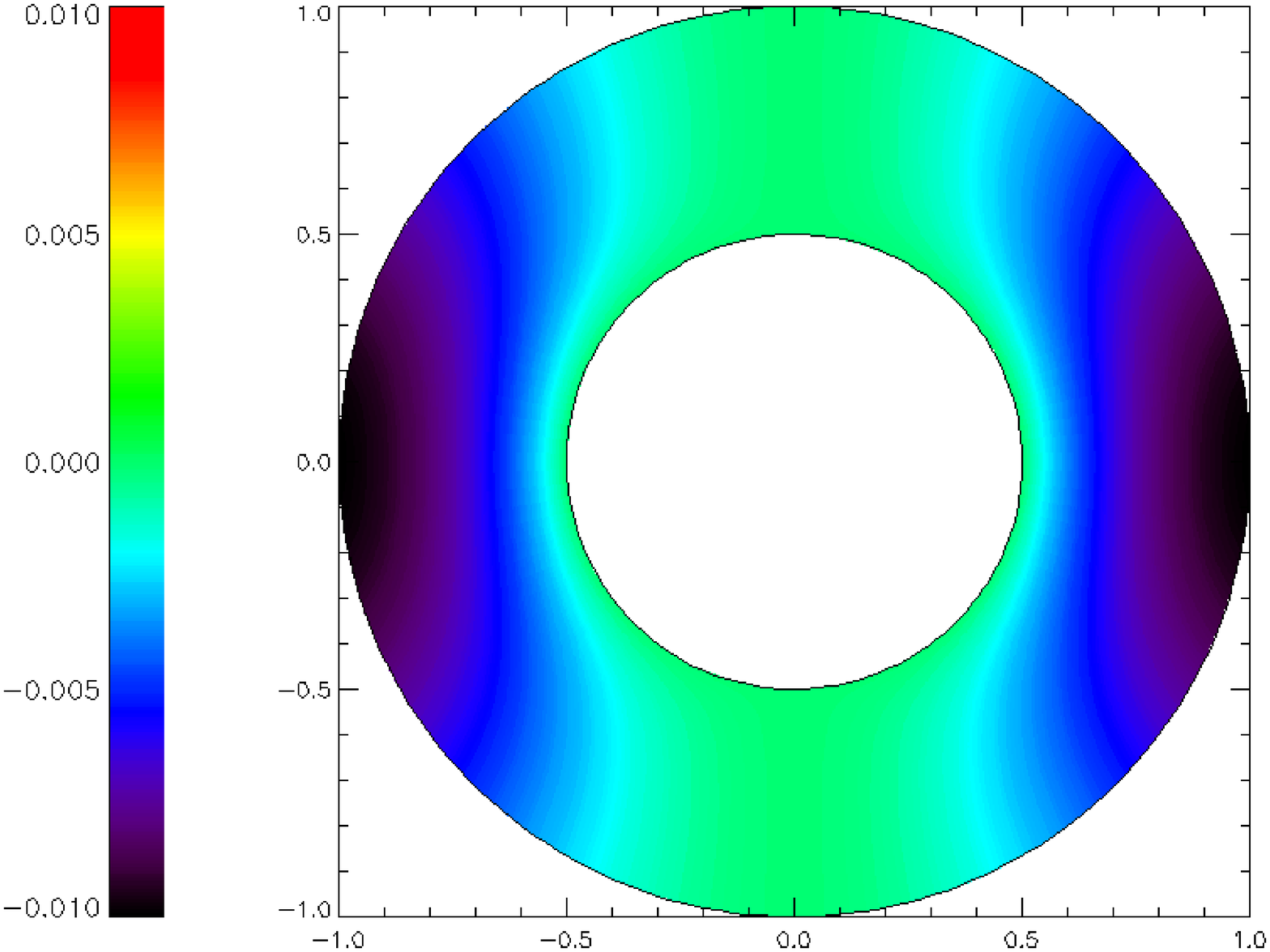}}
    \end{tabular}
    \caption{Response of the fluid due to the boundary forcing with $A=10^{-2}$ for a non-rotating body. Top: Radial velocity in the equatorial plane. Bottom: Radial velocity in the meridional plane corresponding the dotted white line in the top figure. With our scaling, $\textrm{max}(u_r)=1$ when $A=1$. The dominant azimuthal mode $m=2$ and meridional mode $l=2$ are clearly visible. As time evolves, this pattern rotates around the vertical axis at the frequency $\omega$, following the motion of the companion\label{fig:resp}.}
  \end{center}
\end{figure}

The outer boundary condition is used to drive the flow.
Here we choose to impose a radial velocity on a rigid spherical boundary at $r=r_{e}$, which represents radial motions associated with rising and falling of the equilibrium tide at the surface of the body.
This model is valid in the limit of frequencies low compared to surface gravity wave frequencies and small Ekman number, as demonstrated for the linear problem in \cite{ogilvie2009}.
In the case of a tidal potential dominated by the $l=m=2$ spherical harmonic, the radial velocity at the outer boundary can be written as 
\begin{eqnarray}
\label{eq:forcing}
u_r(r=r_e)=\sqrt{\frac{32\pi}{15}}A \Re \left\{Y_2^2(\theta,\phi)e^{-\textrm{i}\omega t}\right\},
\end{eqnarray} 
where $A$ is an arbitrary (real) amplitude.
Using this definition, the maximum radial velocity at the outer boundary is equal to unity when $A=1$.
Note that this choice of boundary condition is compatible with the incompressibility condition, as there is no net mass flux through the boundary.

In reality, as the angular momentum of the system evolves, the amplitude of the tidal forcing will also vary with time (as the fluid synchronises its spin with the tidal deformation, the radial displacement decreases). Our model does not capture this effect. However, since we do not intend to evolve the system until synchronism is reached (the spin typically evolves by only a few per-cent in our simulations), the differences between our results and those for which this effect is taken into account should be minor (we have verified that this is the case for a couple of example simulations in which this effect was taken into account.).

If the companion has an eccentric or inclined orbit, other spherical harmonics than the one present in equation \eqref{eq:forcing} can be excited at various frequencies (e.g.~\citealt{ogilvie2004,BarkerOgilvie2009}).
However, here we focus on the particular component of the tidal potential as defined by equation~\eqref{eq:forcing}, since it is usually the dominant one.
In particular, this is the sole relevant component for the synchronisation problem, in which a companion on a circular orbit in the equatorial plane of the primary orbits with an orbital frequency that is not synchronous with the spin frequency of the fluid, so that $\omega \neq 0$.

At the inner boundary, we neglect the elasticity of the core, and treat it as a rigid spherical boundary on which an impenetrability condition is imposed, so that $u_r(r=r_i)=0$.
The radius of the inner core strongly influences the excitation of inertial waves (e.g.~\citealt{GoodmanLackner2009,Ogilvie2013}).
We did not change the geometry of the core in this paper, in order to focus on studying the dominant nonlinear effects.
For numerical simplicity, we assume that both the inner core and outer boundary are stress-free. We therefore have $\bm{e}_r\times\bm{\sigma}\bm{e}_r=0$ at $r=r_i$ and $r=r_e$, where $\bm{\sigma}=\nu\left(\nabla\bm{u}+\left(\nabla\bm{u}\right)^T\right)$ is the stress tensor and $\bm{e}_r$ is the unit vector in the radial direction.
This is a good approximation for the outer boundary, but is only a simplification for the inner boundary, which is adopted mainly to avoid numerical constraints due to the generation of Ekman boundary layers. We later discuss the case of a no-slip inner core in section \ref{sec:noslip}, since this is probably a better approximation to the boundary of the solid core of a giant planet.

In the case of a non-rotating body, the response of the fluid to this boundary forcing can be seen in Fig.~\ref{fig:resp}.
In a uniformly rotating body, inertial waves are generated at the critical latitude on the inner core \citep{GoodmanLackner2009,sauret} whenever $-2<\omega/\Omega<2$.

\subsection{Energy and angular momentum \label{sec:angmom}}

To help with the analysis of our simulations, we define the volume-integrated kinetic energy, work done by the tidal forcing, and dissipation rate, by
\begin{eqnarray}
K &=& \frac{1}{2}\int_{V}|\boldsymbol{u}|^{2} \ \textrm{d}V, \\
I &=& -\int_{V}\boldsymbol{u}\cdot \nabla W \ \textrm{d}V, \\
D &=& -\nu \int_{V}\boldsymbol{u}\cdot \nabla^2 \boldsymbol{u} \ \textrm{d}V.
\end{eqnarray}
The kinetic energy evolves in time according to
\begin{eqnarray}
\partial_{t}K = I-D.
\end{eqnarray}
We have checked that this is satisfied in the numerical simulations, indicating that our solutions are numerically converged.
The total angular momentum, defined as
\begin{equation}
\boldsymbol{L} = \int_{V} \boldsymbol{x}\times \boldsymbol{u} \ \textrm{d}V,
\end{equation}
is another important quantity in calculations with stress-free boundaries (e.g. \cite{jones2011}).
It evolves according to
\begin{equation}
\label{eq:angmomev}
\frac{\partial \boldsymbol{L}}{\partial t}=-2\int_{V}\boldsymbol{x}\times(\boldsymbol{\Omega}\times\boldsymbol{u}) \ \textrm{d}V-\int_{V}\boldsymbol{x}\times\left(\boldsymbol{u}\cdot\nabla \boldsymbol{u}\right) \ \textrm{d}V \ .
\end{equation}
Note that with spherical boundaries, the pressure torque vanishes, and with stress-free conditions, so does the viscous torque.
Note also that, if we write $\tilde{\boldsymbol{u}} = \boldsymbol{u} + \boldsymbol{\Omega}\times\boldsymbol{x}$, then the only contribution to the right hand side of equation \eqref{eq:angmomev} is
\begin{equation}
\int_{V} \boldsymbol{x}\times \left(\tilde{\boldsymbol{u}}\cdot\nabla \tilde{\boldsymbol{u}}\right) \ \textrm{d}V = -\int_{\partial V} \left(\boldsymbol{x}\times \tilde{\boldsymbol{u}}\right)\tilde{\boldsymbol{u}}\cdot\boldsymbol{n} \ \textrm{d}S\ne 0 \ ,
\end{equation}
where $\boldsymbol{n}$ is the normal to the outer boundary.
This term is nonzero, in general, because the horizontal components of the velocity are nonzero at the boundaries, and a nonzero $u_{r}$ is imposed, i.e., this leads to an angular momentum flux through the outer boundary.
Hence, this term is responsible for causing secular angular momentum evolution.
This term is nonzero even for a linear problem, as long as $\Omega\ne 0$.
However, for a linear problem, the period-averaged contribution from this term vanishes and there is no net change in the angular momentum.
Note however that linear theories can still be used to calculate the tidal torque, which is a quantity of second order in the tidal amplitude.

For linear calculations with $l=m=2$ forcing, we correctly observe only small amplitude oscillatory behaviour in the components of $\boldsymbol{L}$, with zero mean.
However, in the nonlinear simulations, we observe net growth in vertical angular momentum $|L_{z}|$, as is consistent with our outer boundary condition, and represents the process of tidal synchronisation of the spin and orbit.
We do not observe $|L_{x}|$ and $|L_{y}|$ to grow appreciably, which shows that numerical errors are playing a negligible role in the simulations (simulations with no-slip boundary conditions have an oscillatory Ekman layer, which exhibits oscillatory behaviour in $L_{x},L_{y}$ with zero mean, which we present in section \ref{sec:noslip}).

\subsection{Numerical methods \label{sec:num}}

In this paper we use two different numerical approaches to solving equations \eqref{eq:momentum}-\eqref{eq:incomp}, which we will now briefly describe.
Since the two numerical methods are very different, a careful comparison is required, in particular for the nonlinear solutions where previously published results are unavailable.

\subsubsection{PARODY}

In the following, we adapt and use the code PARODY\footnote{\url{http://www.ipgp.fr/~aubert/DMFI.html}} to solve Eqs.~\eqref{eq:momentum}--\eqref{eq:incomp}.
This code was originally written by E. Dormy \citep{dormy1998} and later improved by J. Aubert \citep{aubert2008}.
PARODY has been benchmarked against other numerical codes in the context of a convectively-driven dynamo problem \citep{christensen2001}.
The code is parallelised using both OpenMP and MPI. The time stepping is achieved using a mix of semi-implicit Crank-Nicholson scheme for the linear terms and a second-order Adams-Bashforth scheme for the nonlinear terms.

\begin{table}
 \centering
  \caption{Typical resolutions used within PARODY in the linear regime. $N_R$ is the radial resolution whereas $N_L$ is the maximum degree of the Legendre polynomials. For the linear calculations we only include one azimuthal mode number, $m=2$.\label{tab:one}}
  \begin{tabular}{@{}ccc@{}}
  \hline
   Ekman number & $N_R$ & $N_L$ \\
 \hline
   $10^{-4}$ & $240$ & $32$ \\
   $10^{-6}$ & $960$ & $256$ \\
   $10^{-8}$ & $3840$ & $1024$ \\
 \hline
\end{tabular}
\end{table}

The velocity field is written using a poloidal-toroidal decomposition, thus ensuring incompressibility, with
\begin{equation}
\label{eq:poltor}
\bm{u}=\nabla\times\nabla\times\left(S\bm{e}_r\right)+\nabla\times\left(T\bm{e}_r\right) \ ,
\end{equation}
where $T$ is the toroidal component and $S$ is the poloidal component, and $\bm{e}_r$ is the unit vector in the radial direction.
Each of these scalars is decomposed onto spherical harmonics,
\begin{equation}
S=\sum S_l^m(r)Y_l^m(\theta,\phi) \quad \textrm{and} \quad T=\sum T_l^m(r)Y_l^m(\theta,\phi) \ ,
\end{equation}
where the sum is carried over integers such that $0 \le m \le l$. 
The radial functions $T_l^m(r)$ and $S_l^m(r)$ are represented by their discretized values on a non-uniform radial grid between the inner core located at $r=r_i$ and the outer core located at $r=r_e$. The radial derivatives are computed using second order finite-differences. The grid is denser close to the inner and outer boundaries in order to appropriately resolve flows near the boundaries.
This mesh refinement at the boundaries is particularly important when using no-slip boundary conditions in order to accurately resolve thin Ekman boundary layers.

Due to the poloidal-toroidal decomposition \eqref{eq:poltor}, the implementation of a non-vanishing radial velocity at one of the boundaries is non-trivial.
The stress-free condition imposes
\begin{equation}
\frac{\partial \left(T_l^m/r\right)}{\partial r}=0 \quad \textrm{(instead of} \ T_l^m=0 \ \textrm{in the no-slip case)}
\end{equation}
for all $l,m$ at both $r=r_i$ and $r=r_e$.
We focus here on a radial forcing corresponding to $l=2$ and $m=2$ as defined by equation \eqref{eq:forcing}.
The boundary condition for the poloidal component at the outer radius $r_e$ is in this case 
\begin{equation}
\frac{\partial^2 S_l^m}{\partial r^2}=
    -A\frac{16}{r_e^2}\sqrt{\frac{2\pi}{15}}S_l^m \quad \text{if $l=m=2$},
\end{equation}
or zero otherwise, and
\begin{equation}
S_l^m=    A\frac23\sqrt{\frac{2\pi}{15}}r_ee^{\textrm{i}\omega t} \quad \text{if $l=m=2$},
\end{equation}
or zero otherwise. The inner core is stress-free and impenetrable, with
\begin{equation}
\frac{\partial^2 S_l^m}{\partial r^2}=S_l^m=0
\end{equation}
for all $l,m$ at $r=r_i$.

The main advantage of PARODY is the spectral decomposition in the azimuthal direction, which allows us to focus on a limited range of azimuthal modes.
In the linear case, one can take advantage of the fact that only modes with $m=2$ are excited, effectively reducing the problem to two dimensions.
The numerical convergence of the code has been checked for various Ekman numbers and forcing frequencies, and the typical resolution in the meridional plane used for production runs is shown in Table \ref{tab:one}.
In the nonlinear case, we also checked the numerical convergence with the number of azimuthal modes and we typically use $m_{max}=64$ (which is sufficient for amplitudes $A\lesssim 10^{-2}$).

\subsubsection{Nek5000}

For comparison with the nonlinear solutions obtained using PARODY, we also use the efficiently parallelised spectral element code Nek5000, written by Paul Fischer and collaborators \citep{Fischer2007,nek5000}.
Spectral element methods combine the geometric flexibility of finite element methods with the accuracy of spectral methods, and solve the weak variational form of the equations of motion, similarly to finite element methods \citep{Fischer2002}.
This method partitions the domain into a set of $\mathcal{E}$ non-overlapping sub-domains, called elements, whose union is the entire domain.
Within each element the velocity components and the pressure are represented as tensor product Lagrange interpolation polynomials of order $N$ and $N-2$, respectively, defined at the Gauss-Lobatto-Legendre and Gauss-Legendre points.
Such a method has algebraic convergence with increasing $\mathcal{E}$, but spectral (exponential) convergence with increasing $N$ (for smooth solutions), with the total number of grid points in 3D being $\mathcal{E}N^3$.

Temporal discretisation is based on a semi-implicit formulation, where the nonlinear and Coriolis terms are treated explicitly and the viscous terms are treated implicitly.
In particular, a 3rd order backward-difference formula is used for the viscous \& pressure terms, and a 3rd order extrapolation is used for the explicit terms.
Dealiasing is used, with the polynomial order $N$ increased by a factor of 3/2 for the evaluation of non-linear (advective) terms. In our spherical shell computations, the points lie on spherical shells, to double precision. Typical resolutions adopted are $\mathcal{E}=576$ and $N=20$ (30 for the nonlinear terms). 

For the computations reported in this paper, PARODY tends to be somewhat more efficient, primarily due its ability to simulate a restricted range of azimuthal mode numbers $m$, which is found to be sufficient when $A\ll1$. However, Nek5000 is likely to be more efficient at very high resolution, since the Legendre transform requires global communication of all spectral coefficients, which is not required in spectral element methods. Both codes have been compared for both linear and nonlinear problems, as we will describe below.

%
%
\begin{figure}
  \begin{center}
    \begin{tabular}{c}
      \hspace{4mm}\resizebox{65mm}{!}{\includegraphics{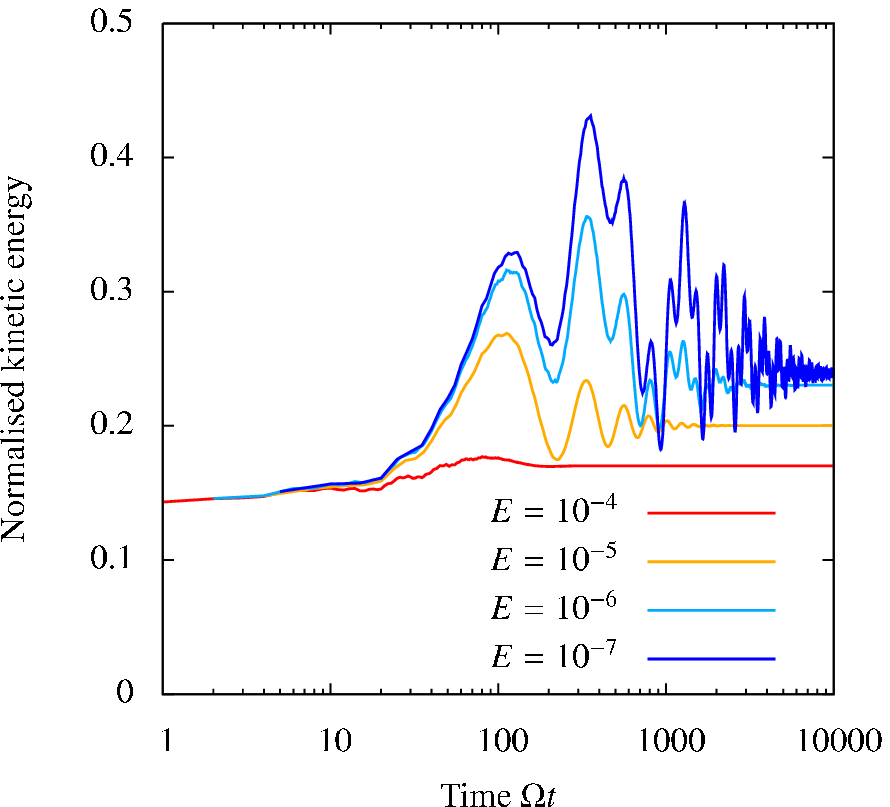}} \\
      \resizebox{70mm}{!}{\includegraphics{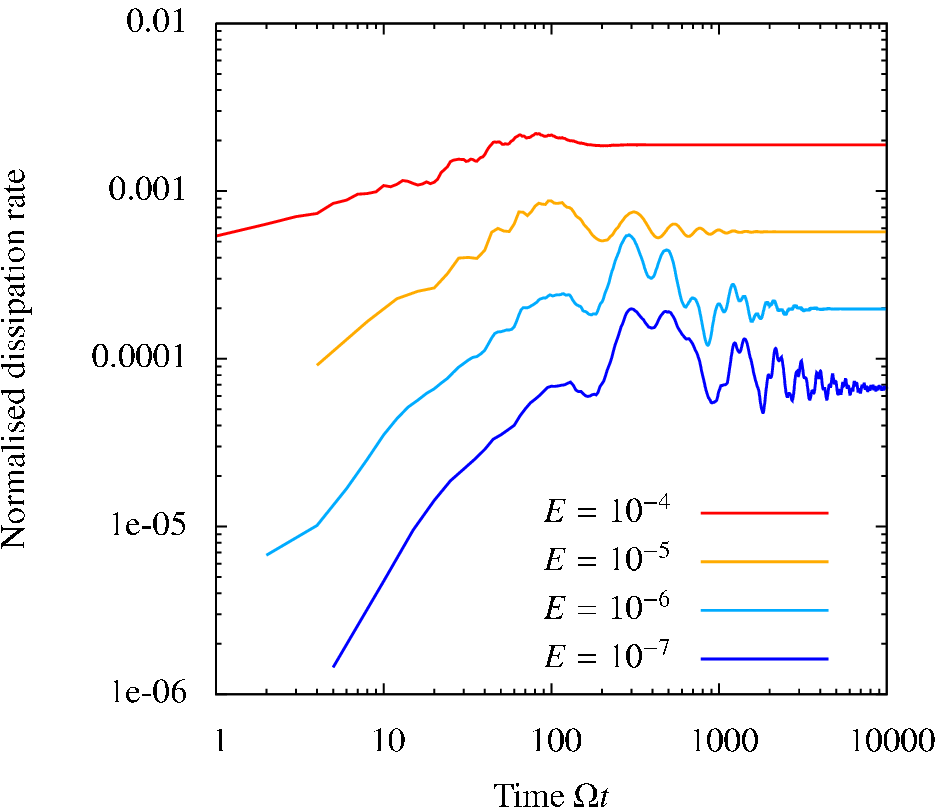}}
    \end{tabular}
    \caption{Kinetic energy (in units of $\rho r_e^3A^2$) and dissipation rate (in units of $\rho r_e^3 A^2 \Omega$) versus time for $\omega/\Omega=\sqrt{2}$ and various Ekman numbers. The time is expressed in units of the spin frequency $\Omega$.\label{fig:ex}}
  \end{center}
\end{figure}
%
%
%
\begin{figure*}
  \begin{center}
    \begin{tabular}{c}
      \hspace{-8mm}
      \resizebox{48mm}{!}{\includegraphics{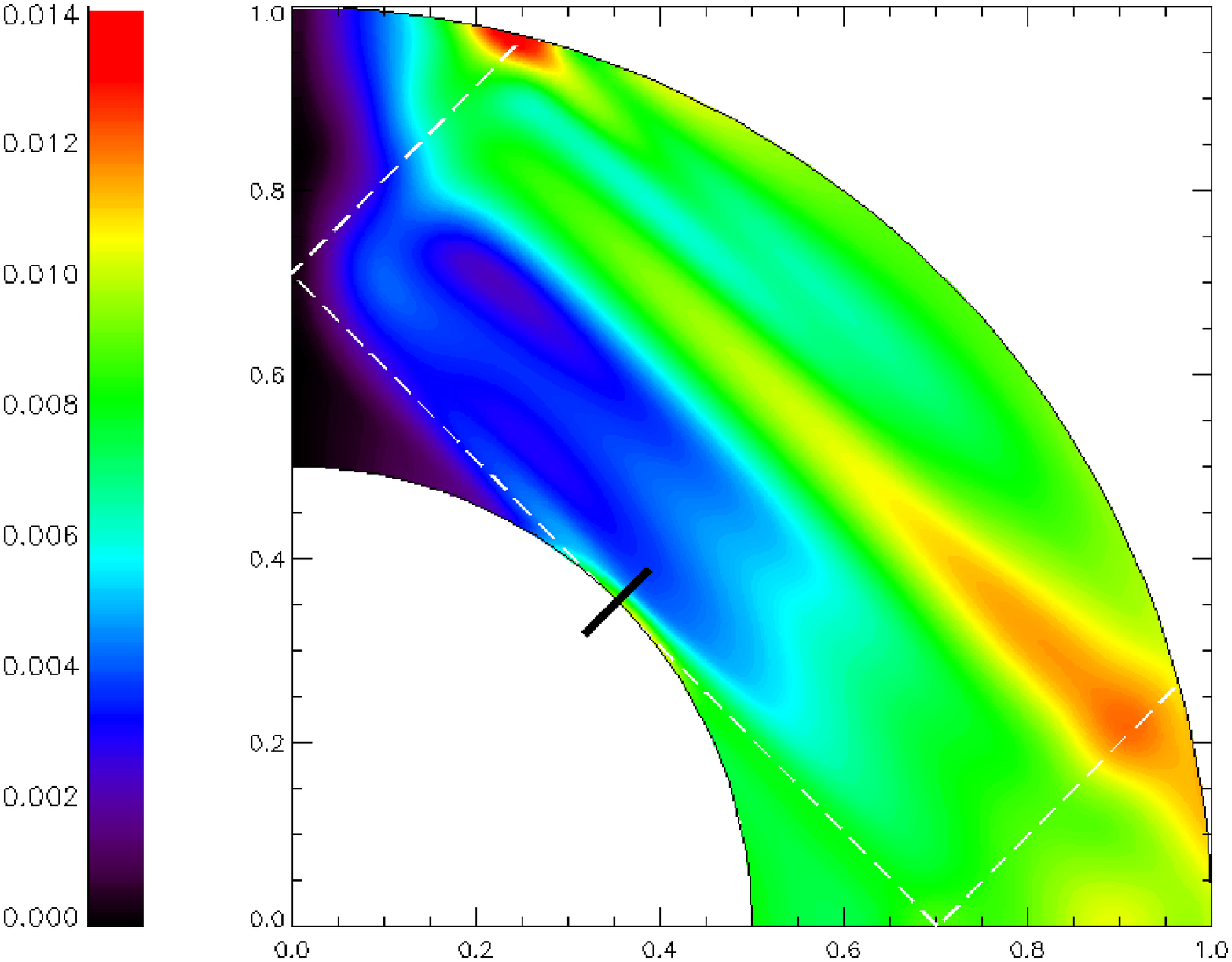}}
      \resizebox{48mm}{!}{\includegraphics{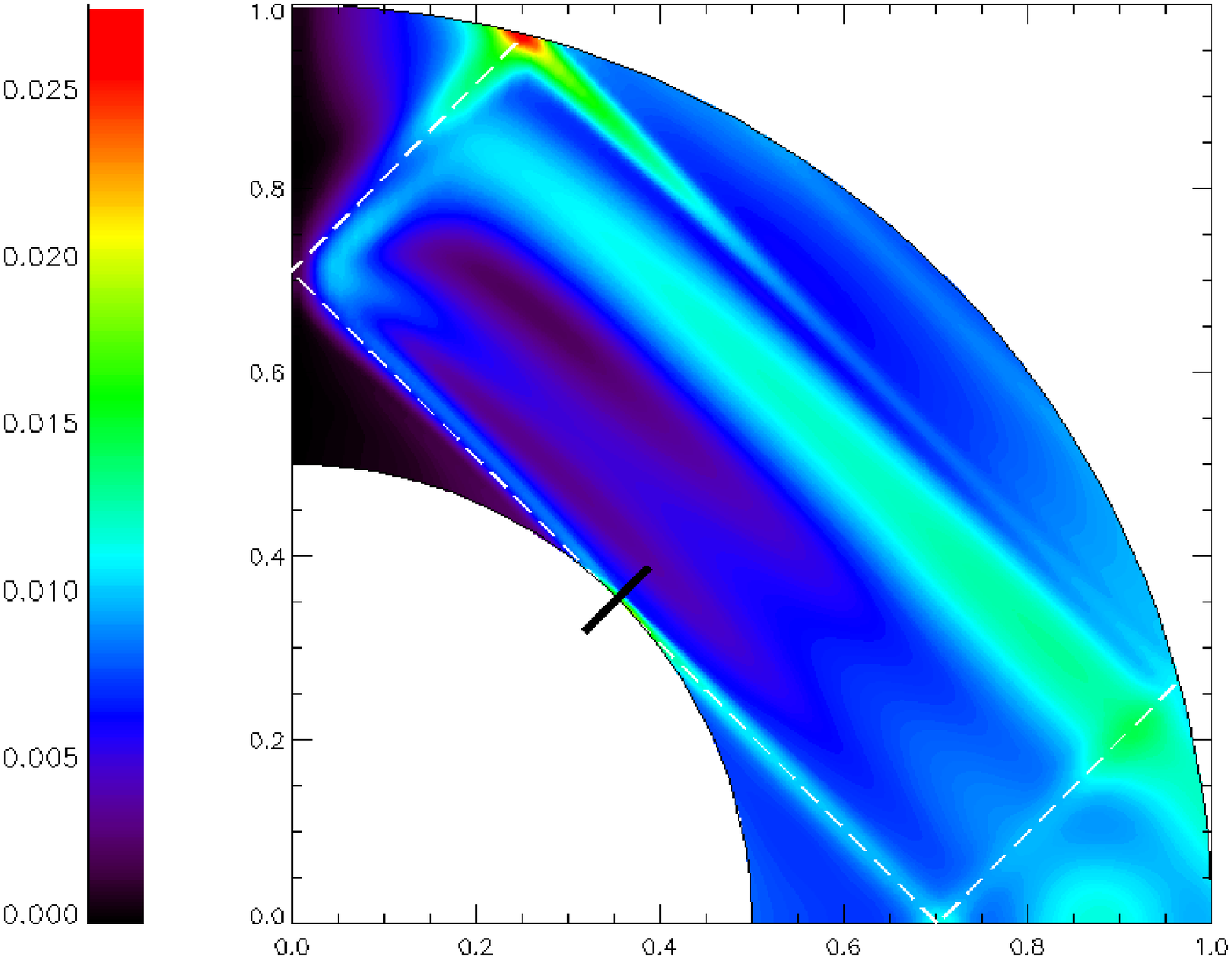}}
      \resizebox{48mm}{!}{\includegraphics{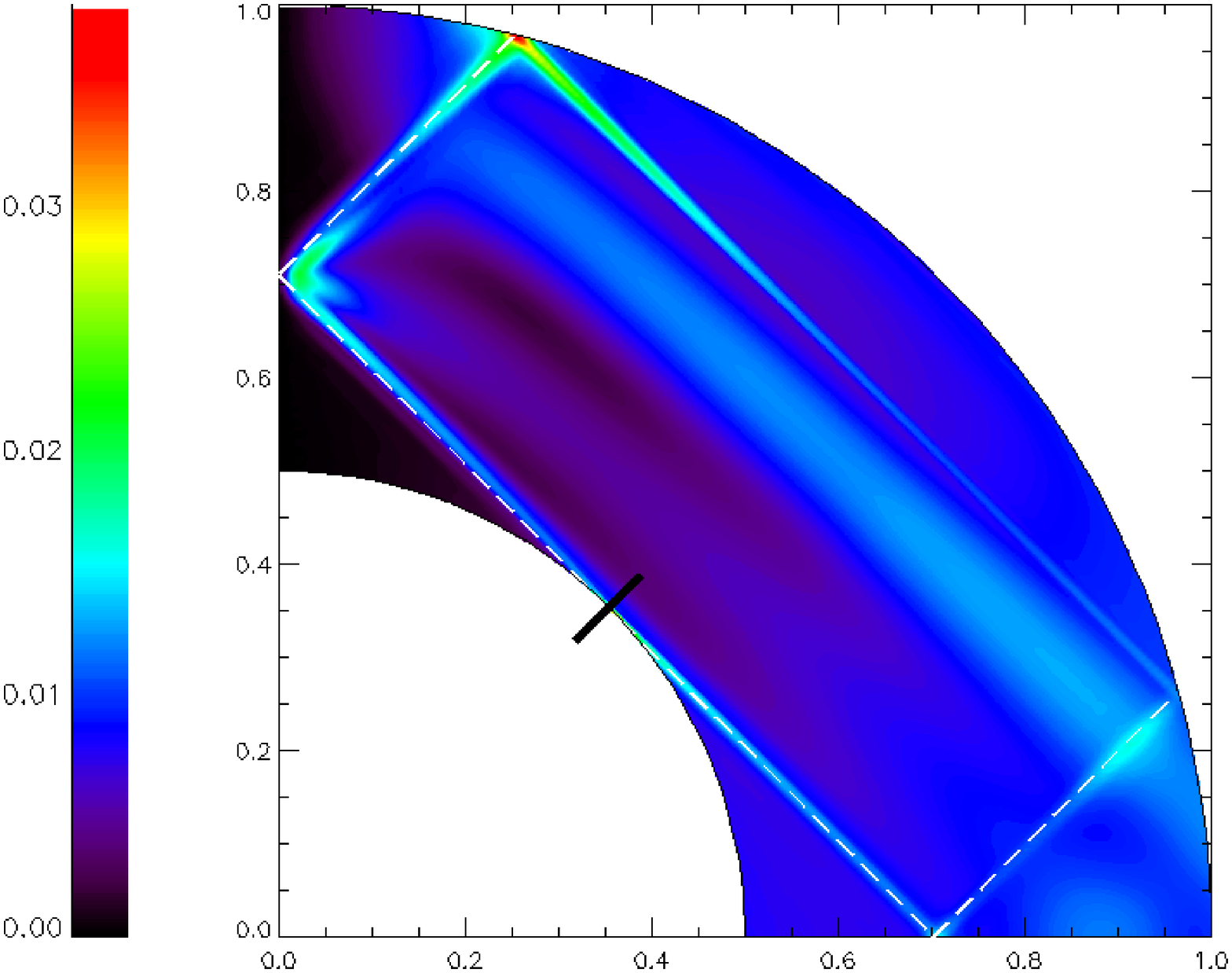}}
      \resizebox{35mm}{!}{\includegraphics{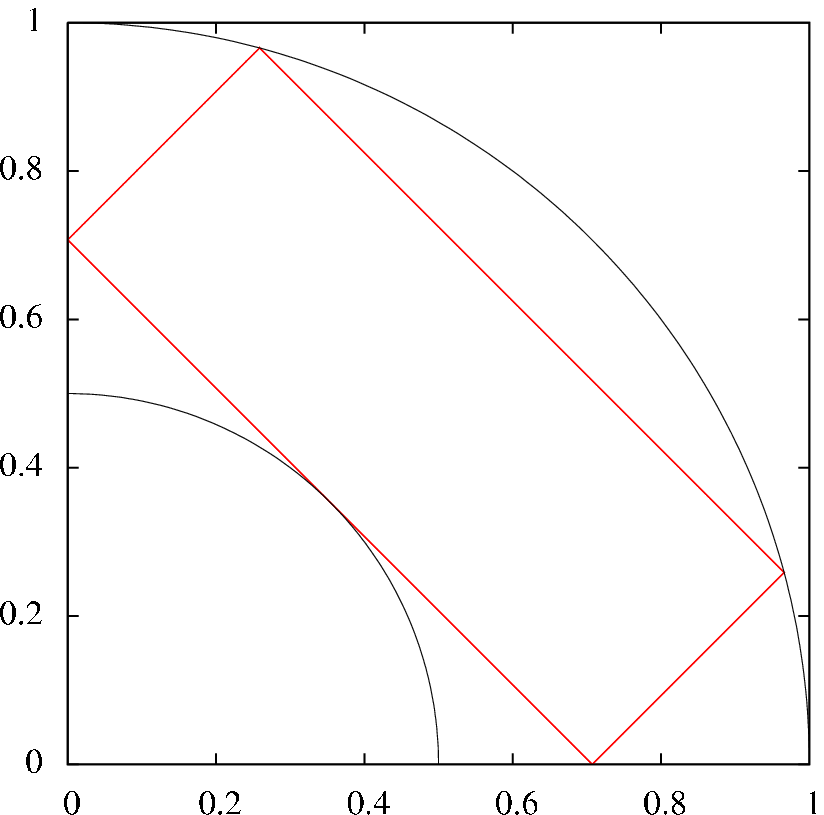}}
    \end{tabular}
    \begin{picture}(10,0)
        \put(-170,92){\scriptsize{$E=10^{-5}$}}
        \put(-31,92){\scriptsize{$E=10^{-6}$}}
        \put(108,92){\scriptsize{$E=10^{-7}$}}
    \end{picture}
    \caption{Azimuthally-averaged velocity magnitude $|\bm{u}|$ is the meridional plane.
The solution is symmetric with respect to the equator so that only the upper quadrant is shown.
In all cases, black corresponds to zero whereas red corresponds to the largest value.
The frequency is $\omega/\Omega=\sqrt{2}$ and the Ekman number is $E=10^{-5}$, $10^{-6}$ and $10^{-7}$ moving from left to right in the panels.
The black straight line shows the location of the critical latitude defined by the polar angle $\theta_c=\textrm{arccos}(\omega/2)$ at the inner core.
The dotted white line is tangent to the inner core and corresponds to the direction of propagation of the inertial waves emitted at the critical latitude on the inner core.
On the right, we show the path of characteristics initialised at the inner critical latitude.\label{fig:u_ex}}
  \end{center}
\end{figure*}
\section{Linear regime}\label{sec:linear}

In this section, we illustrate the model described in section \ref{sec:model} and compare some of our results with previous works in the linear regime.
This is to illustrate both the transient phase before our solution converges to a steady state, and to check that our predictions for the steady state in the linear regime are consistent with previously published results.
Numerically, the linear regime is recovered by effectively switching off the second term on the left-hand side of equation \eqref{eq:momentum} and fixing the amplitude $A$ to an arbitrary value equal to unity.
In this section, we primarily use PARODY, since this allows us to focus on the linear response of the $m=2$ modes.
PARODY and Nek5000 have been found to agree accurately for the linear problem for all frequencies compared when $E=10^{-5}$.
Nek5000 was not used to explore smaller Ekman numbers because the mesh adopted is fully three-dimensional, whereas PARODY can exploit the $m=2$ symmetry, which reduces the dimensionality of the linear problem.
The typical resolution required for such linear simulations can be found in Table~\ref{tab:one}.

\subsection{Illustration of the model\label{sec:ex}}

We first illustrate how this model behaves for a particular frequency $\omega/\Omega=\sqrt{2}$.
This frequency was considered by \cite{rieutord2010}, because the path of characteristics generated at the critical latitude converges towards a simple rectangular shape (see the right panel in Fig.~\ref{fig:u_ex}).
The group velocity of an inertial wave is proportional to its wavelength and is inclined at an angle $\textrm{arcsin}(\omega/2\Omega)$ to the rotation axis, which is the angle at which the (inviscid) rays propagate.
The critical latitude is the location where the inertial waves propagate tangentially to the boundary.
We compute the response of the fluid for four different Ekman numbers from $E=10^{-4}$ down to $E=10^{-7}$.

We plot in Fig.~\ref{fig:ex} the total kinetic energy and the total dissipation rate versus time.
In all cases, the system reaches a steady-state after a significant fraction of a global viscous timescale.
Note that while we reach a steady state for $E=10^{-4}$, $E=10^{-5}$ and $E=10^{-6}$, transients are still present at the end of our $E=10^{-7}$ simulation.
For this particular frequency, the dissipation rate decreases (not linearly) as the Ekman number is decreased.
As we will see in the next section, this is not the case for all frequencies.
As already discussed in section \ref{sec:angmom}, there is no net evolution in the vertical component of the angular momentum.
We show the azimuthally-averaged velocity magnitude $|\bm{u}|$ in the meridional plane in Fig.~\ref{fig:u_ex}.
This flow pattern corresponds to the steady state in each case.

As the Ekman number decreases, it becomes particularly clear that an inertial wave beam is emitted at the critical circle at the inner boundary defined by $\theta=\textrm{arccos}(\omega/2\Omega)$ and $r=r_i$.
The emergence of waves from the critical circle has been noticed before \citep{tilgner1999,ogilvie2004,ogilvie2009}.
In the inviscid limit, the flow is singular there \citep{stewartson1969} and the solution is regularised by viscosity.
The path of characteristics emerging from the inner critical latitude in the inviscid limit is shown in the rightmost panel in Fig.~\ref{fig:u_ex}.

\subsection{Comparison with previous works}\label{sec:comp_gordon}

\begin{figure}
  \begin{center}
    \begin{tabular}{c}
      \hspace{-0.6cm}\resizebox{80mm}{!}{\includegraphics{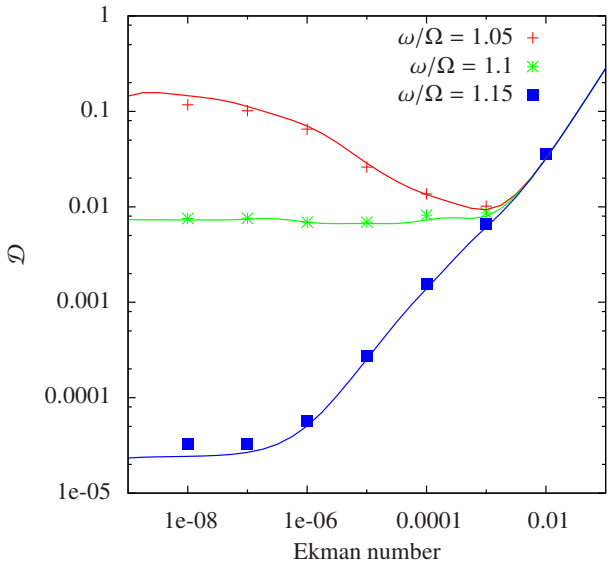}}
    \end{tabular}
    \caption{Dissipation rate (in units of $\rho r_e^3 A^2 \Omega$) for $\omega/\Omega=1.05$, $1.1$ and $1.15$ and various Ekman numbers.
Lines are reproduced from the Fig.7 of Ogilvie (2009) whereas symbols correspond the steady-state of our initial value problem solved with PARODY.
\label{fig:compgordon}}
  \end{center}
\end{figure}

In this section, we compare our results with \cite{ogilvie2009}.
Three close frequencies are compared: $\omega/\Omega=1.05$, $\omega/\Omega=1.1$ and $\omega/\Omega=1.15$.
While these frequencies are very similar, the dissipation rate was shown to crucially depend on the Ekman number in a very different way for each of these frequencies.
More specifically, the dissipation was roughly independent of the Ekman number for $\omega/\Omega=1.1$, whereas it increases as the Ekman number is decreased for $\omega/\Omega=1.05$, with the opposite behaviour for $\omega/\Omega=1.15$.
We reproduce this result here using our initial value approach.
The Ekman number is varied from $E=10^{-2}$ down to $E=10^{-8}$ and the aspect ratio of the spherical shell is again $r_i/r_e=0.5$.
The simulations are run until a steady state is reached, and we then measure the corresponding dissipation rates.
We show in Fig.~\ref{fig:compgordon} the results reproduced from \cite{ogilvie2009} superimposed with the results from our initial value approach.
The agreement with the results from PARODY is excellent, with slight differences at the smallest Ekman number of $10^{-8}$, for which the high resolution used (see table \ref{tab:one}) implies millions of iterations before reaching a steady-state.
This comparison is to be considered as a consistency check only, as our initial value problem is much more numerically demanding than the associated direct linear calculation of a steady-state response.

\subsection{Scalings with the Ekman number: nonlinearities in the astrophysical regime}
\begin{figure}
  \begin{center}
    \begin{tabular}{c}
      \resizebox{70mm}{!}{\includegraphics{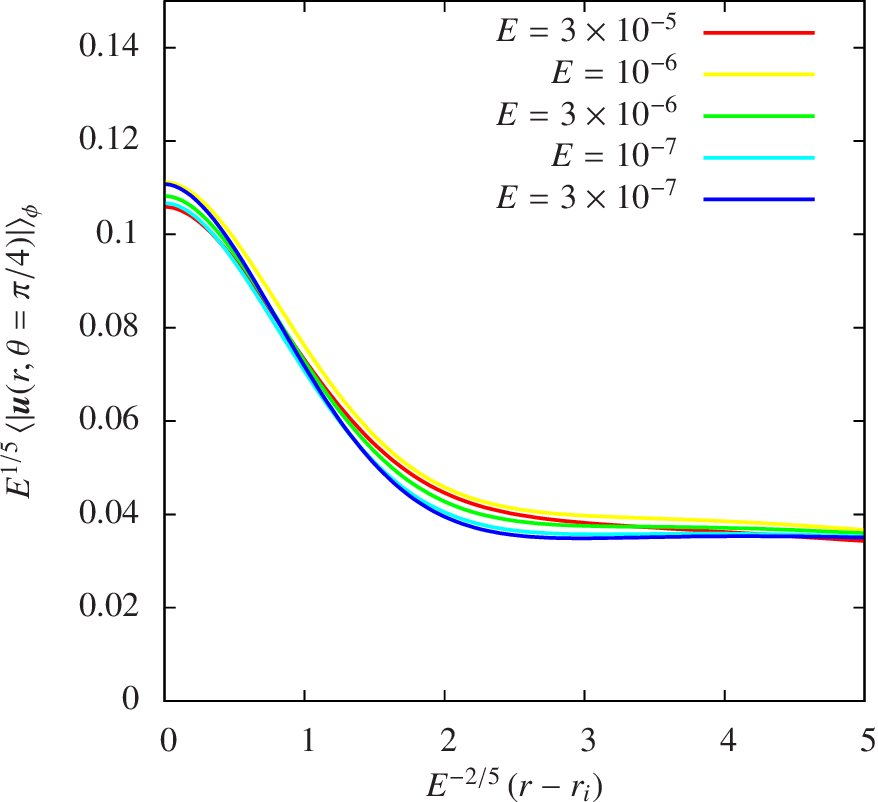}}\\
      \resizebox{70mm}{!}{\includegraphics{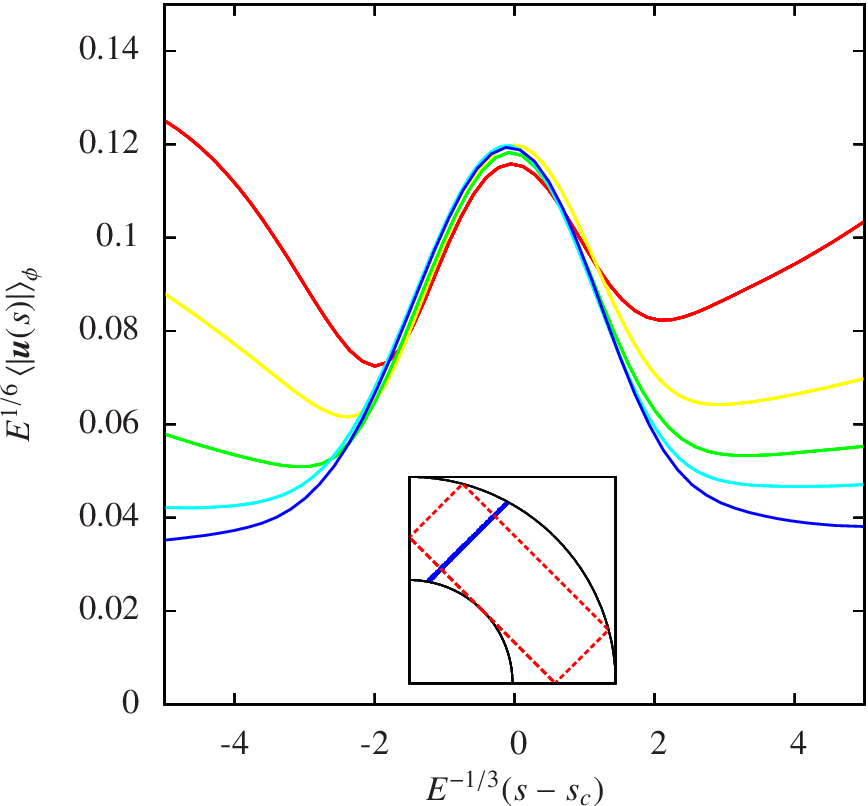}}
    \end{tabular}
    \caption{Scalings with the Ekman number $E$.
Top: Azimuthally-averaged velocity amplitude along radius at fixed polar angle $\theta=\pi/4$ which corresponds to the critical latitude.
We focus on the region close to $r=r_i$ which corresponds to the critical latitude for $\omega/\Omega=\sqrt{2}$.
The radius is normalised by $E^{2/5}$ whereas the amplitude is normalised by $E^{-1/5}$.
Bottom: Azimuthally-averaged velocity amplitude along the blue line shown in the embedded panel.
The spatial coordinate along this line $s$ is centred around the second intersection with the path of characteristics shown in the embedded figure and normalised by $E^{1/3}$.
The amplitude is normalised by $E^{-1/6}$.\label{fig:scalings}}
  \end{center}
\end{figure}
\cite{stewartson1972} showed that the singularities associated with the solutions to the Poincar\'e equation should turn into oscillating shear layers through viscosity.
The width of these regions depends on the mechanism that generates them and must scale with the Ekman number.
Finding scalings of this kind is important because it allows us to extrapolate our results to the astrophysically relevant regime, in which the Ekman number takes extremely small values
Most of the previous analytical studies were derived in the context of no-slip boundaries so that Ekman boundary layers scaling as $E^{1/2}$ are present, which are not expected in our problem with stress-free boundaries.
However, we can numerically determine the scalings from the results of the linear calculations presented in Section~\ref{sec:ex} for $\omega/\Omega=\sqrt{2}$.
From Fig.~\ref{fig:u_ex}, it is clear that the width of the shear layers scales as a positive power of the Ekman number.
In order to quantify this scaling, we plot in Fig.~\ref{fig:scalings} the azimuthally averaged velocity amplitude along two lines inclined with an angle $\textrm{arcsin}(\omega/2\Omega)=\pi/4$ with respect to the vertical axis.
One of the lines intersects the origin $r=0$ and passes through the inner critical latitude whereas the second one intersects the $z$-axis at $z=0.4$ (see embedded plot in Fig.~\ref{fig:scalings}).
In each of these plots, the spatial coordinate and the amplitude are compensated by some power of the Ekman number.

As expected from previous analysis \citep{kerswell1995}, the radial width of the shear layer at the critical latitude on the inner boundary scales as $E^{2/5}$ while the width of the internal shear layers in the direction normal to the path of characteristics scales as $E^{1/3}$.
Note that $E^{1/4}$ is another expected scaling but our results do not allow us to distinguish between $E^{1/3}$ and $E^{1/4}$ \citep{rieutord2001}.
The amplitude of the flow at the critical latitude and in the shear layers is another important issue.
The amplitude of the velocity at the critical latitude scales as $E^{-1/5}$ whereas the amplitude of the velocity inside the shear layers scales as $E^{-1/6}$ (see the scalings used in Fig.~\ref{fig:scalings}).
Note that we could not check the dependence of these scalings on the Doppler-shifted frequency $\omega$.
We only manage to unambiguously determine the scalings for $\omega/\Omega=\sqrt{2}$ because the path of characteristics is spatially very simple in that particular case (see Fig.~\ref{fig:u_ex}).
For other frequencies, the proximity and intersection between different wave beams make the scaling analysis difficult, especially at the Ekman numbers available numerically.

We may use the scalings determined above to predict at what values of the input parameters, and at what location in the flow, nonlinearities are likely to become important first.
To do this, we can compare the wave velocity amplitude $u$ with the phase velocity of the local wave packet $\omega/k $, where $k\sim l^{-1}$ is the perpendicular wave number, and $l$ is the width of the wave beam, and suppose that nonlinear terms become important when $\frac{u k}{\omega} \gtrsim 1$.
This can be thought of as a dimensionless measure of the nonlinearity in the wave beams, and it is very likely that an inertial wave beam with an amplitude larger than this will become unstable and break (see \citealt{clark2010,Scolan2013} for inertial gravity waves and \citealt{bordes2012,jouve2013} for inertial waves).
Similar estimates are also appropriate for internal gravity waves (e.g. \citealt{BarkerOgilvie2010,Bourget2013}). 

Using the scalings determined above, the nonlinearity scales as $E^{-3/5}$ at the critical latitude and $E^{-1/2}$, at most, in the internal shear layers/wave beams (taking the thinnest $E^{1/3}$ scaling).
This suggests that in the astrophysical regime, in which $E\rightarrow 0$, the nonlinearity in the vicinity of the critical latitude is probably the most important.
We therefore expect this location to be the one with the dominant nonlinear interactions.
Inertial waves launched from this location might undergo instabilities and break before they can reflect from the boundaries.
If this occurs, this will very likely modify the frequency dependence of the dissipation rate \citep{GoodmanLackner2009}.
However, this might be difficult to capture numerically, since these instabilities probably have much smaller scales than the primary wave beams, therefore these instabilities might be difficult to capture in our simulations. Nevertheless, the aforementioned scalings give some insight into when non-linearities should become prevalent.
From now on, we simulate the effects of these nonlinearities directly.

%
%
\section{Nonlinear regime}\label{sec:nonlinear}

We now move on to the main focus of this work, which is to study the effects of nonlinearities.
The nonlinear term in equation \eqref{eq:momentum} is now taken into account.
This leads to additional numerical constraints.
In particular, it is necessary to include a range of azimuthal wave numbers in PARODY, since the $m=2$ symmetry is no longer preserved in the presence of nonlinear couplings.
The typical resolution in the meridional plane must be also increased for the same reason.
In addition, the Courant-Friedrichs-Lewy stability constraint requires a smaller time step.
Consequently, it is not possible to numerically reach the very low Ekman numbers that were obtainable in the purely linear regime.
Most of the nonlinear simulations presented in this section correspond to $E \ge 10^{-5}$.
The next subsections are devoted to the effect of nonlinearities while varying some of the relevant parameters. Our aim is to determine how nonlinear effects modify the solution from the predictions of linear theory.

\begin{figure}
    \begin{tabular}{c}
      \hspace{-3mm}\resizebox{80mm}{!}{\includegraphics{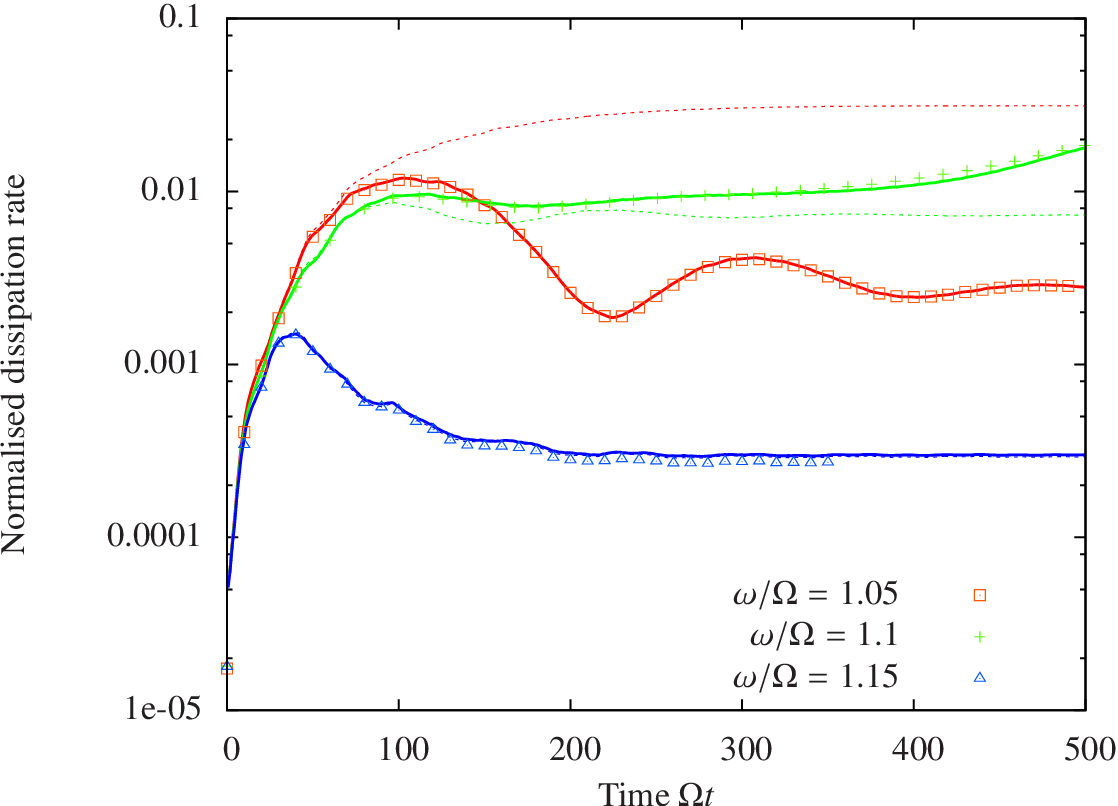}} \\
      \hspace{2mm}\resizebox{75mm}{!}{\includegraphics{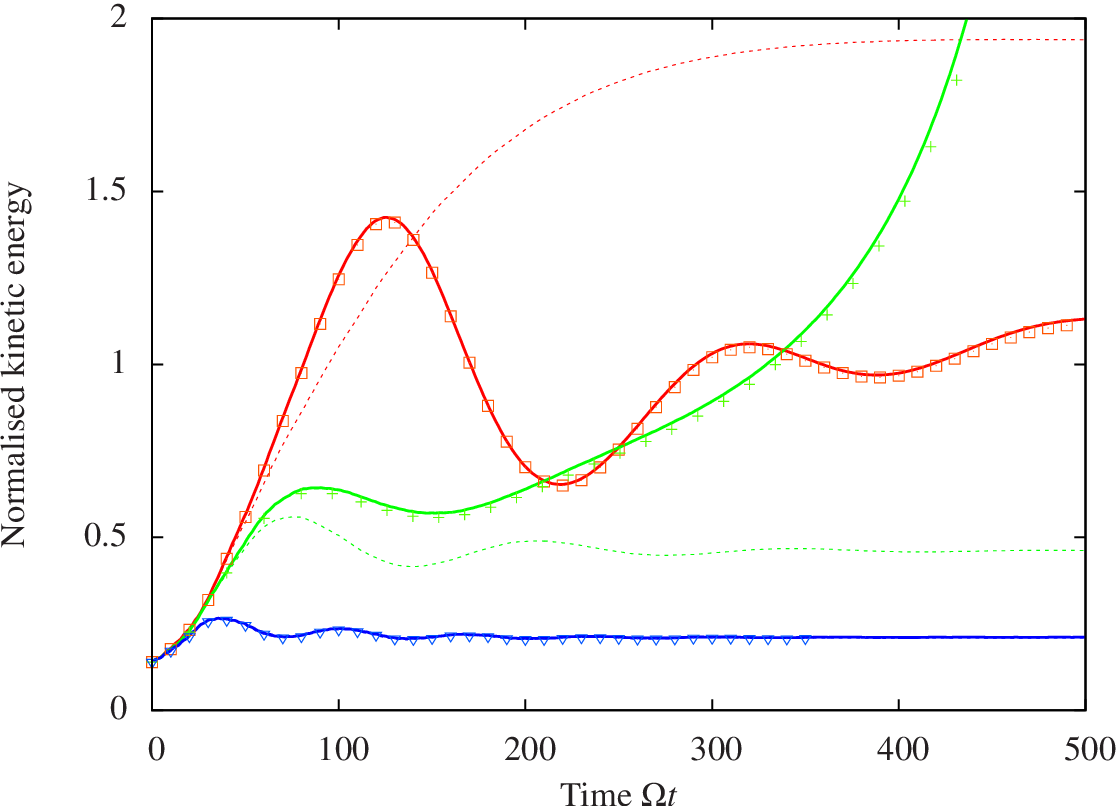}}
    \end{tabular}
  \caption{Comparison between the two codes for a simulation with $A=0.00386$ and $E=10^{-5}$ for the three frequencies $1.05$, $1.10$ and $1.15$.
The solid lines correspond to the results obtained with PARODY whereas the symbols correspond to the results obtained with Nek5000.
The viscous dissipation rate $D$ is shown in units of $r_e^3 A^2 \Omega$ whereas the kinetic energy $K$ is shown in units of $r_e^3 A^2$.
The thin dotted lines correspond to the linear results already presented in Fig.~\ref{fig:compgordon}
\label{fig:3freqcomp}}
\end{figure}
\subsection{Illustration for three frequencies and comparison of the two codes}

To validate the results of our numerical simulations in the nonlinear regime, we compare the results using both codes, for the following simulations.
The Ekman number is fixed to be $E=10^{-5}$ whereas the amplitude of the forcing is $A=3.86 \times 10^{-3}$.
We compare the three frequencies already discussed in Section~\ref{sec:comp_gordon}, namely $\omega/\Omega=1.05$, $\omega/\Omega=1.1$ and $\omega/\Omega=1.15$.
For PARODY, the resolution used is $32$ Fourier modes in the azimuthal direction, $194$ Legendre polynomials and $480$ grid-points in the radial direction.
For Nek5000, the resolution is $\mathcal{E}=576$ and $N=20$ ($30$ for the nonlinear terms).
Note the advantage of using a spectral method in the azimuthal direction, nonlinear couplings are weak for this particular amplitude, so that a relatively small number of azimuthal modes are required in order to reach numerical convergence.
The total kinetic energy and viscous dissipation rate are plotted versus time in Fig.~\ref{fig:3freqcomp}.
The solid lines correspond to the results obtained with PARODY whereas the symbols correspond to the results obtained with Nek5000.
The thin dotted lines correspond to the previous linear results.
Note that in the case $\omega/\Omega=1.15$, the dissipation is much lower than for the two other frequencies and there are barely any differences between the linear and nonlinear solutions for this amplitude.
The agreement between the two codes is excellent, even after $500$ periods ($O(10^{4})$ time steps), bearing in mind that the two numerical schemes are based on very different approaches\footnote{The most likely source of discrepancy between the two codes is that the incompressibility condition is exactly satisfied in PARODY, as a result of the poloidal-toroidal decomposition of the velocity field. On the other hand, Nek5000 solves a discrete Poisson equation for pressure, which it solves by preconditioned conjugate gradient iteration with a given tolerance, which usually results in errors $\lesssim 10^{-7}$ per time step. However, these errors can accumulate over long duration simulations with $\sim 10^{5}$ time steps, which could explain this minor discrepancy.}.
We can therefore explore the nonlinear behaviour of the model with confidence.
\begin{figure}
  \begin{center}
    \begin{tabular}{c}
      \resizebox{60mm}{!}{\includegraphics{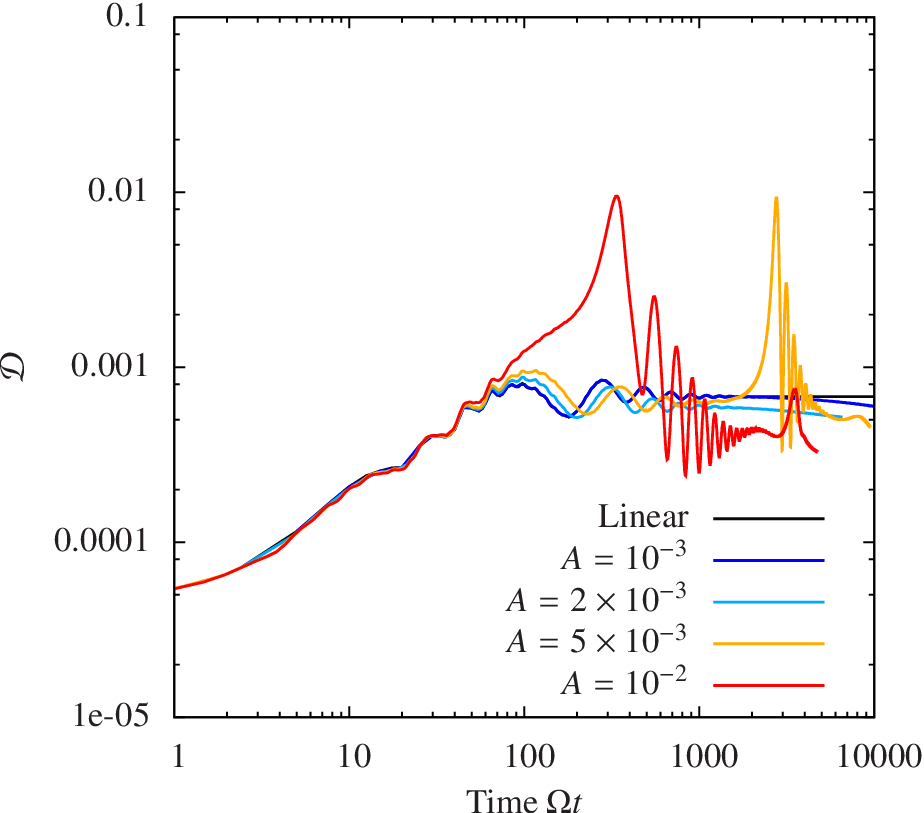}} \\
      \resizebox{60mm}{!}{\includegraphics{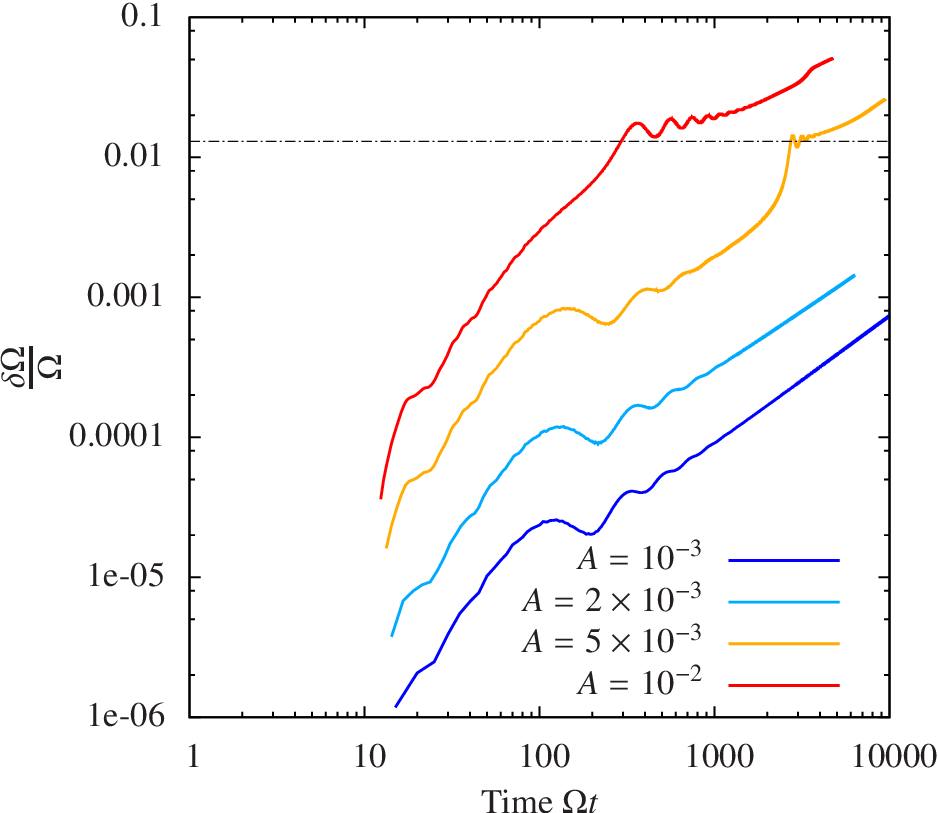}}
    \end{tabular}
    \caption{Dissipation rate (in units of $\rho r_e^3 A^2 \Omega$) and spin frequency of the fluid $\delta\Omega$ in the rotating frame for $\omega/\Omega=\sqrt{2}$ and various amplitudes $A$ of the forcing. The horizontal line shows the transition between two different regimes as soon as $\delta\Omega>10^{-2}$. This transition occurs for amplitudes $A=10^{-2}$ and $A=5\times10^{-2}$ at times $\Omega t\approx300$ and $\Omega t\approx3000$ respectively. Note that during the early evolution of the solution, $\delta \Omega$ scales like $A^{2}$. \label{fig:amp}}
  \end{center}
\end{figure}
\subsection{Increasing the amplitude}
\begin{figure}
  \begin{center}
    \begin{tabular}{c}
      \resizebox{70mm}{!}{\includegraphics{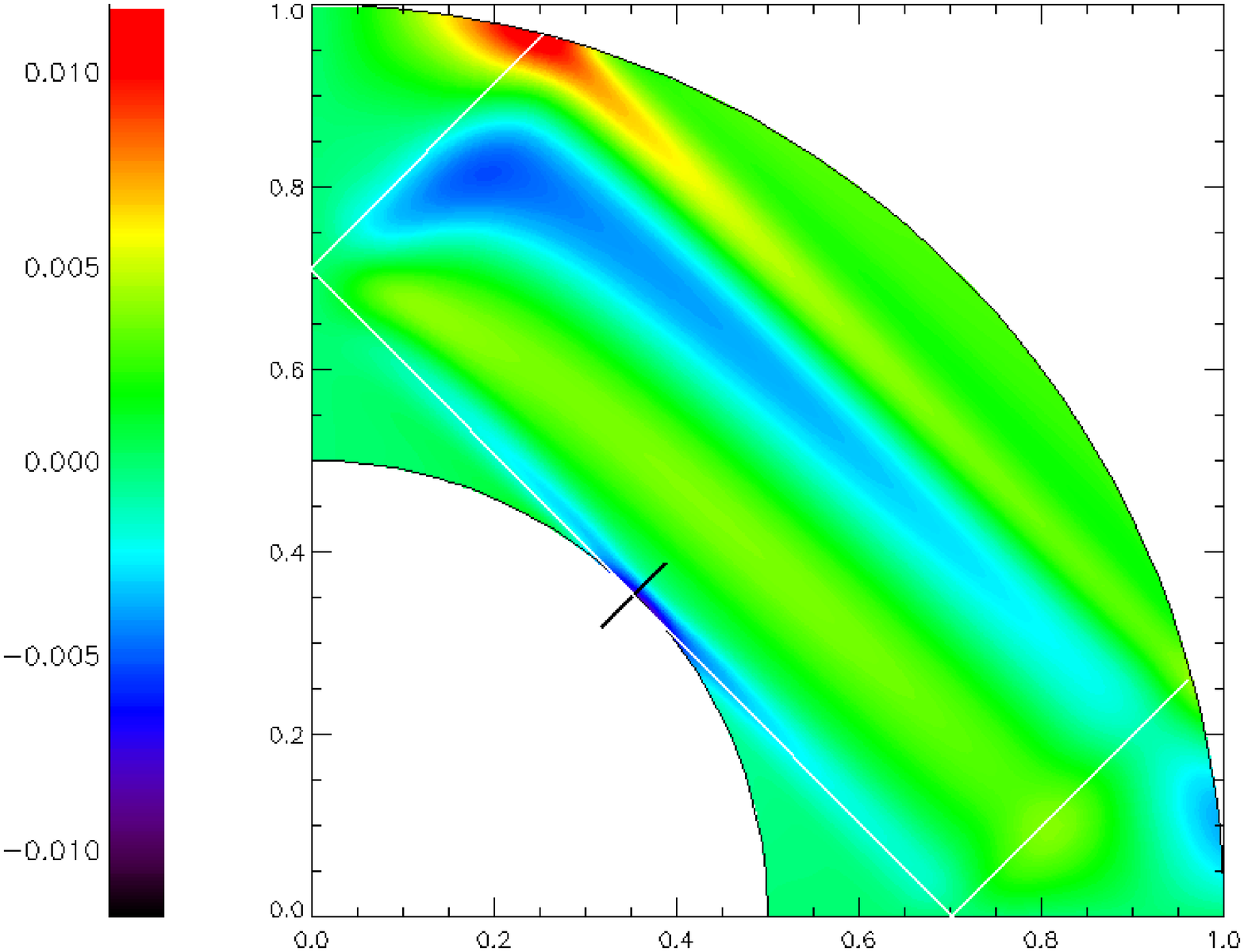}} \\
      \resizebox{70mm}{!}{\includegraphics{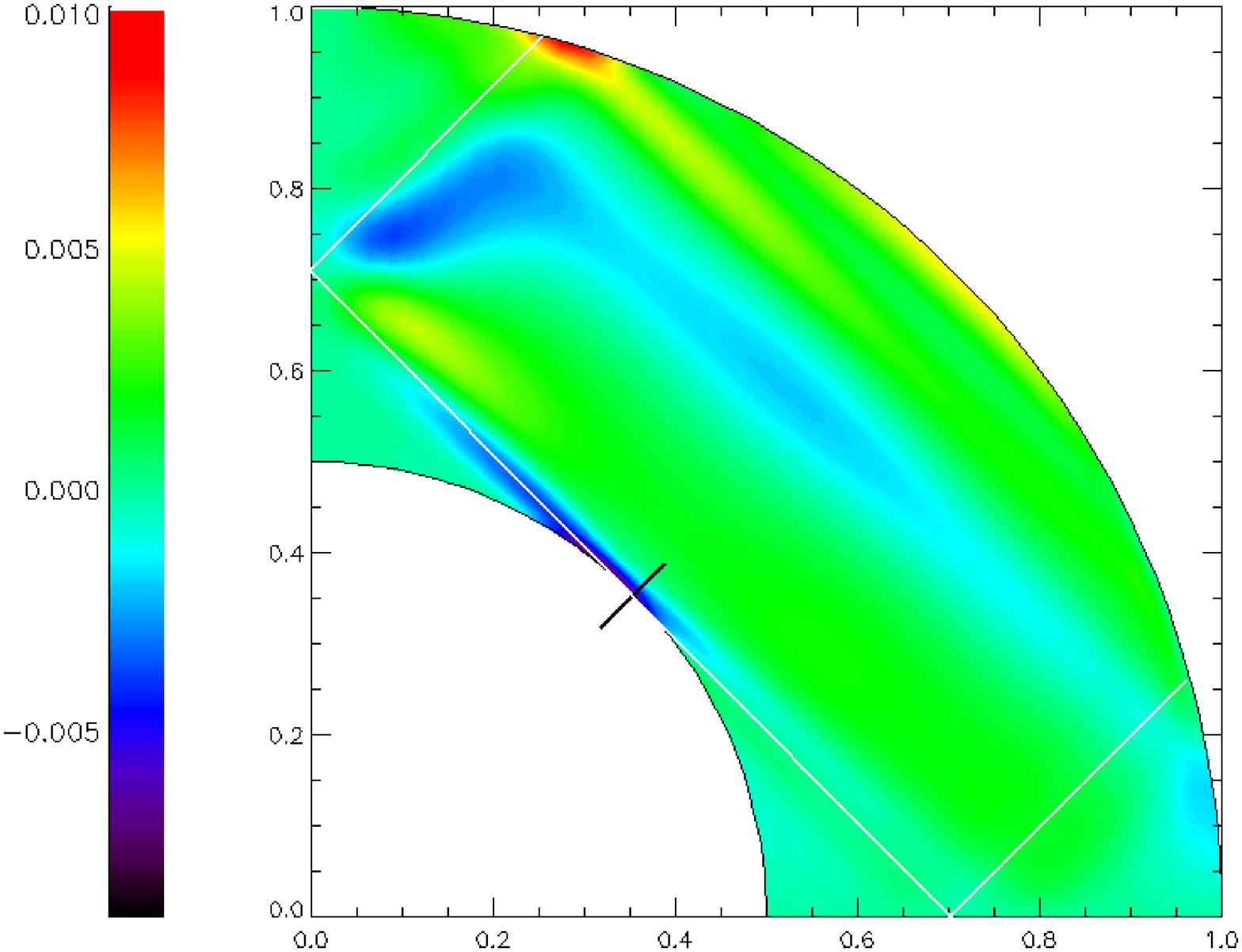}}
    \end{tabular}
    \caption{Normalised tangential velocity $u_{\theta}$ in a meridional slice for $\omega/\Omega=\sqrt{2}$ and $E=10^{-5}$. We compare the linear results (top) during the steady state and the nonlinear regime (bottom) with $A=5\times10^{-3}$ at the arbitrary time $\Omega t \approx  3000$, well into the nonlinear regime.\label{fig:ut}}
  \end{center}
\end{figure}

In this section, we fix $\omega/\Omega=\sqrt{2}$.
This frequency was already considered in the linear regime in Section~\ref{sec:linear}.
We now consider the nonlinear regime by progressively increasing the amplitude of the forcing from $A=10^{-4}$ to $A=10^{-2}$.
The dissipation rate versus time is shown in Fig.~\ref{fig:amp} for the various amplitudes along with the purely linear result.
Contrary to the linear regime, no steady state is reached and the dissipation rate is now a time-dependent quantity, with variations of more than one order of magnitude.
This is primarily because angular momentum is continuously injected through the outer boundary, as we discussed in Section~\ref{sec:angmom}, which is an effect not present in the linear calculations.
This can be seen in Fig.~\ref{fig:amp}, where we plot the increase in the mean rotation rate of the fluid versus time.
In the frame initially rotating at a frequency $\Omega$, we define the increase in the volume-averaged rotation rate of the fluid as
\begin{equation}
\delta\Omega=\frac1V\int\frac{u_{\phi}}{r\sin\theta} \ \textrm{d}V \ ,
\end{equation}
where $u_{\phi}$ is the azimuthal velocity and $V$ is the volume of the spherical shell.
$\delta\Omega$ is initially zero, and evolves with time so that the total rotation rate of the fluid at a given time is $\Omega^*(t)=\Omega+\delta\Omega(t)$.
Contrary to the linear regime, the contributions from the forcing do not average out over one period so that angular momentum is injected into the fluid (if $\omega>0$, extracted otherwise, see Section~\ref{sec:limit}).
This process will continue until the fluid is spinning synchronously with the forcing, \textit{i.e.} when $\omega=\omega_i-2(\Omega+\delta\Omega)=0$.
Even if both the amplitude and the Ekman number in our simulations are typically larger than we would expect in stellar or planetary interiors, it is not possible computationally to run the simulation until complete synchronisation is achieved.

By changing the spin frequency $\Omega+\delta\Omega$ of the fluid, we continuously sweep across different values for the Doppler-shifted forcing frequency $\omega$, so that the properties of the linear excitation of inertial waves varies with time.
It is known from previous linear calculations that the dissipation rate at low Ekman number is a complicated function of the forcing frequency \citep{ogilvie2009}.
By spinning up the fluid, it is therefore unsurprising to observe strong variation in the dissipation rate as time evolves.
In addition, the generation of differential rotation in the bulk i.e. zonal flows, also changes the properties of the inertial wave response \citep{baruteau2013}.
This is discussed further in section \ref{sec:zonal}.
\begin{figure*}
  \begin{center}
    \begin{tabular}{ccc}
      \resizebox{55mm}{!}{\includegraphics{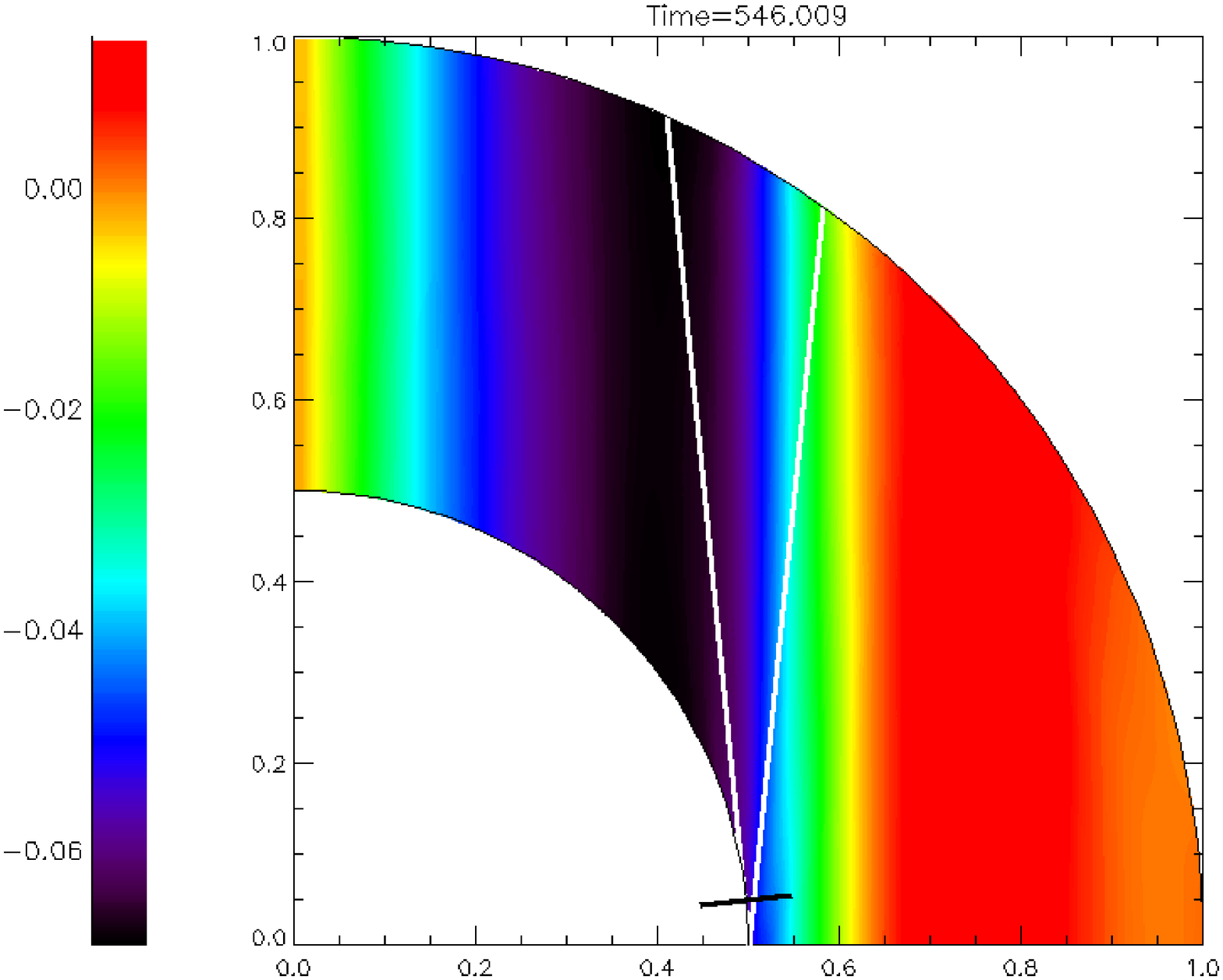}} &
      \resizebox{55mm}{!}{\includegraphics{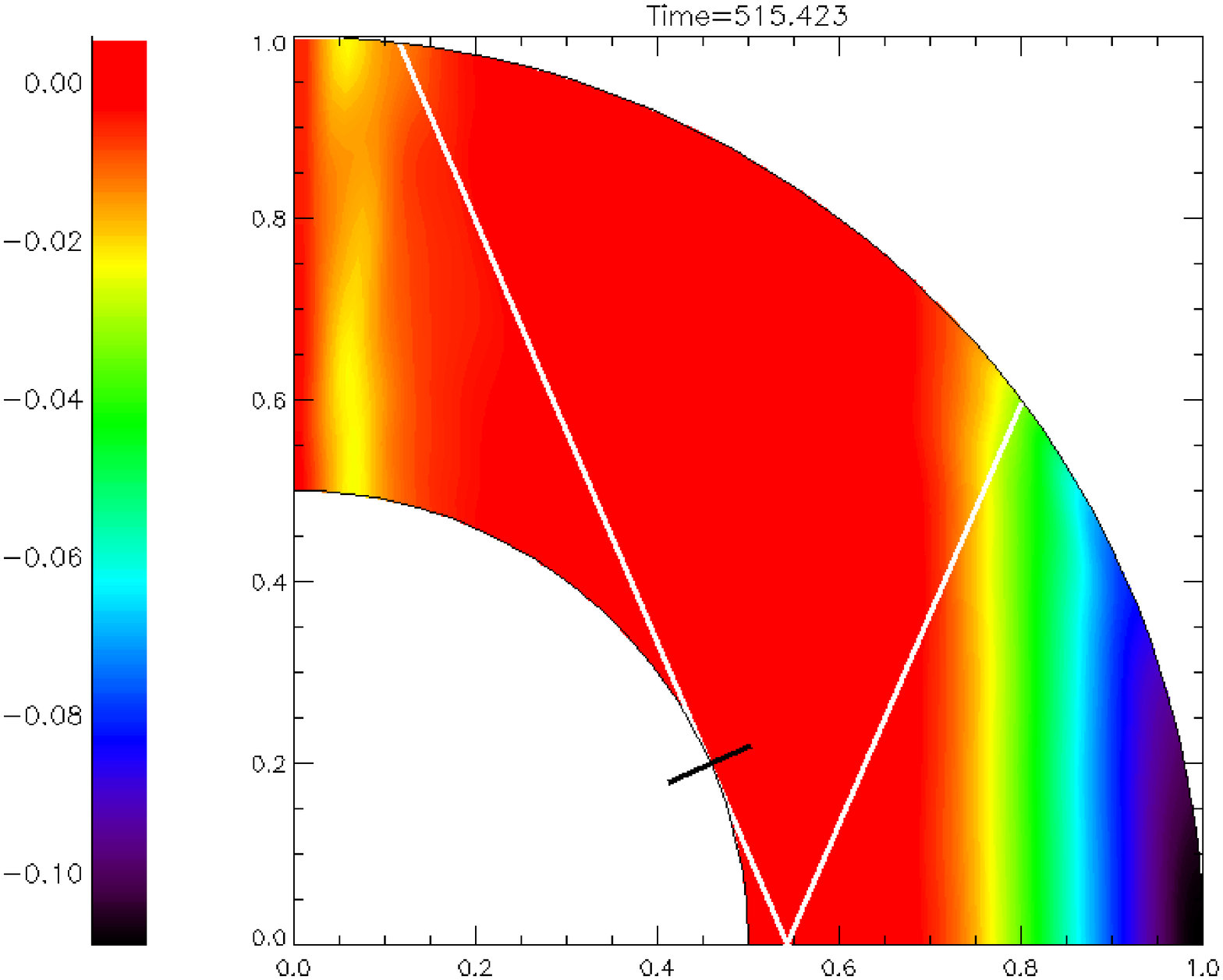}} &
      \resizebox{55mm}{!}{\includegraphics{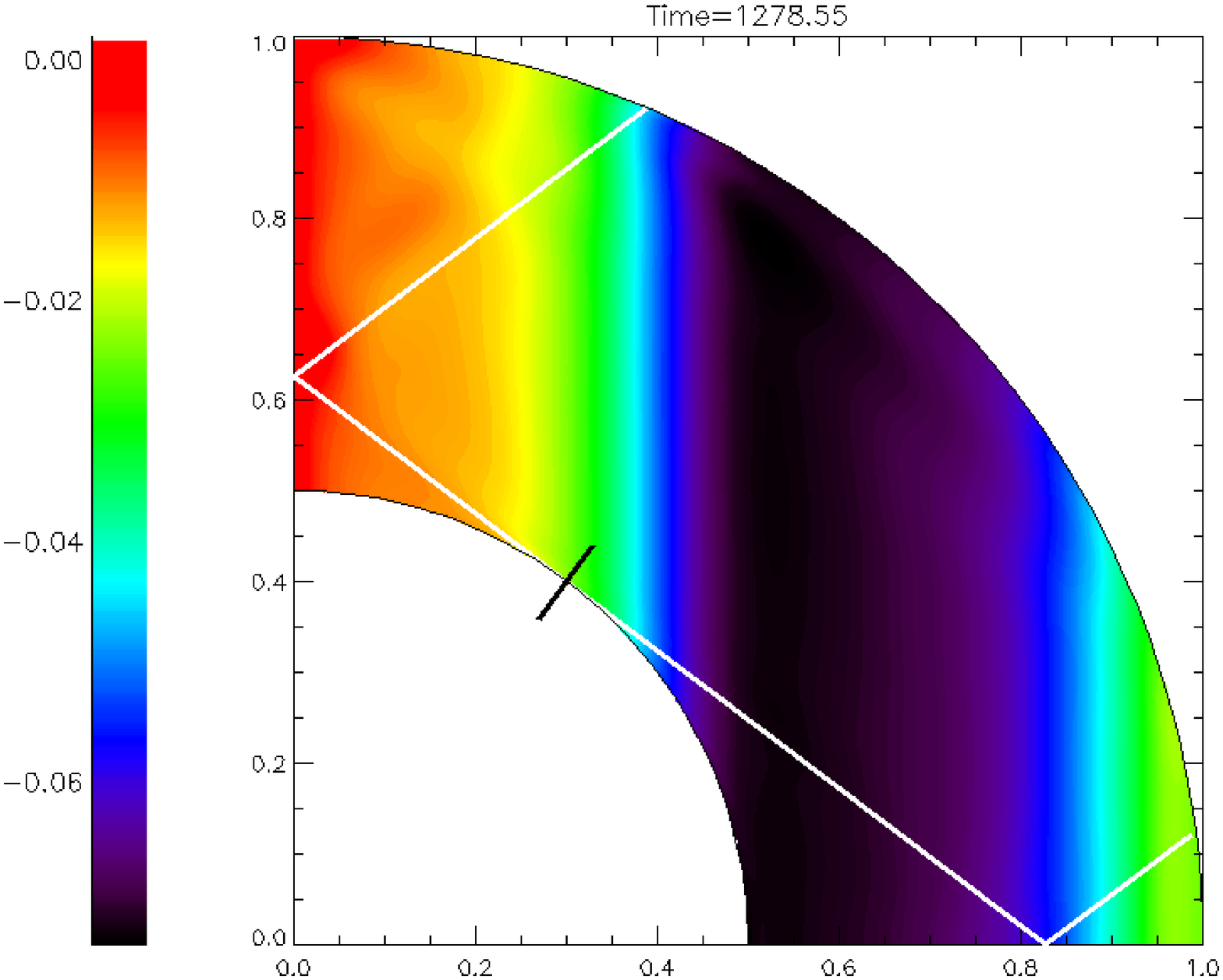}} \\
      \resizebox{55mm}{!}{\includegraphics{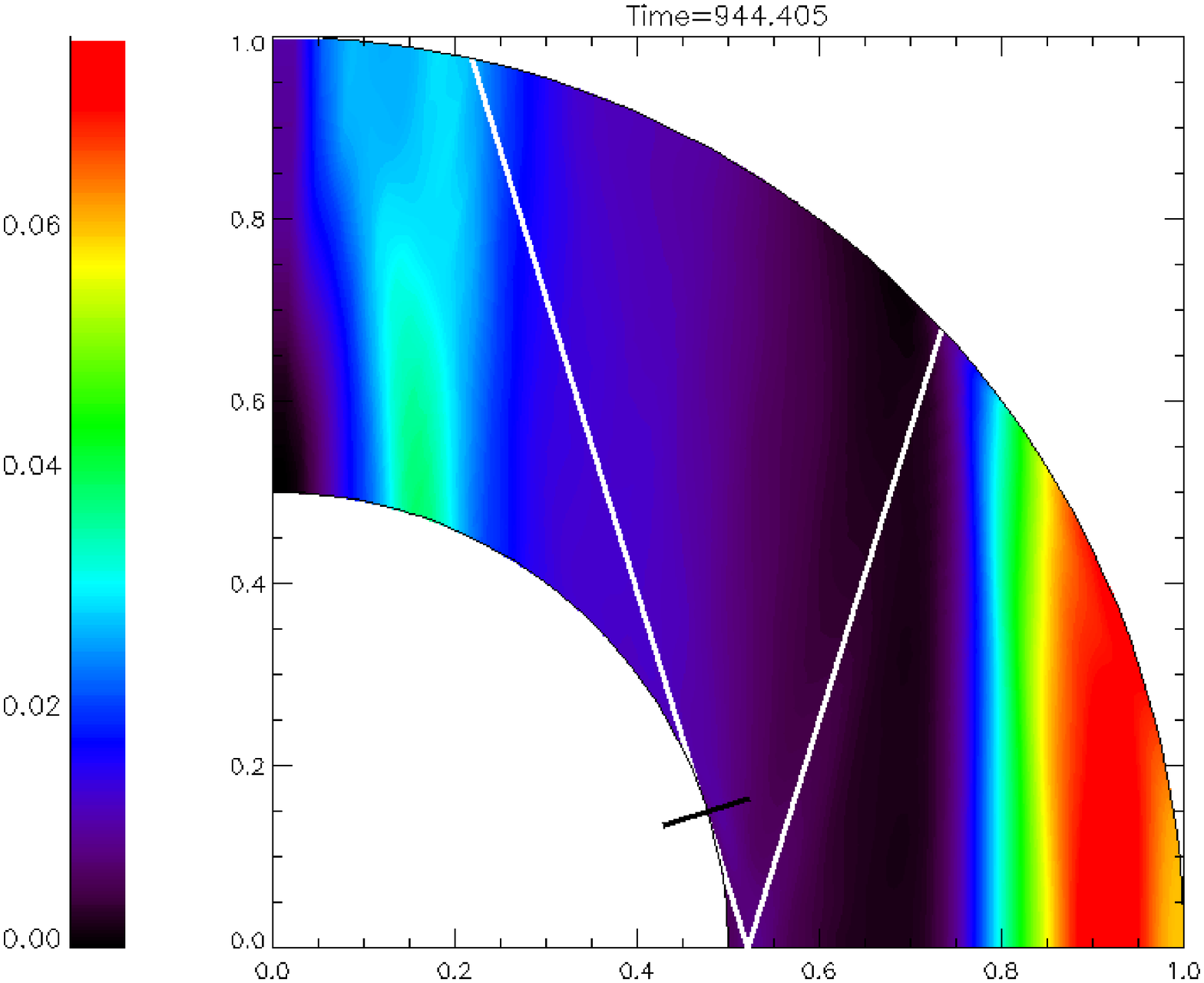}} &
      \resizebox{55mm}{!}{\includegraphics{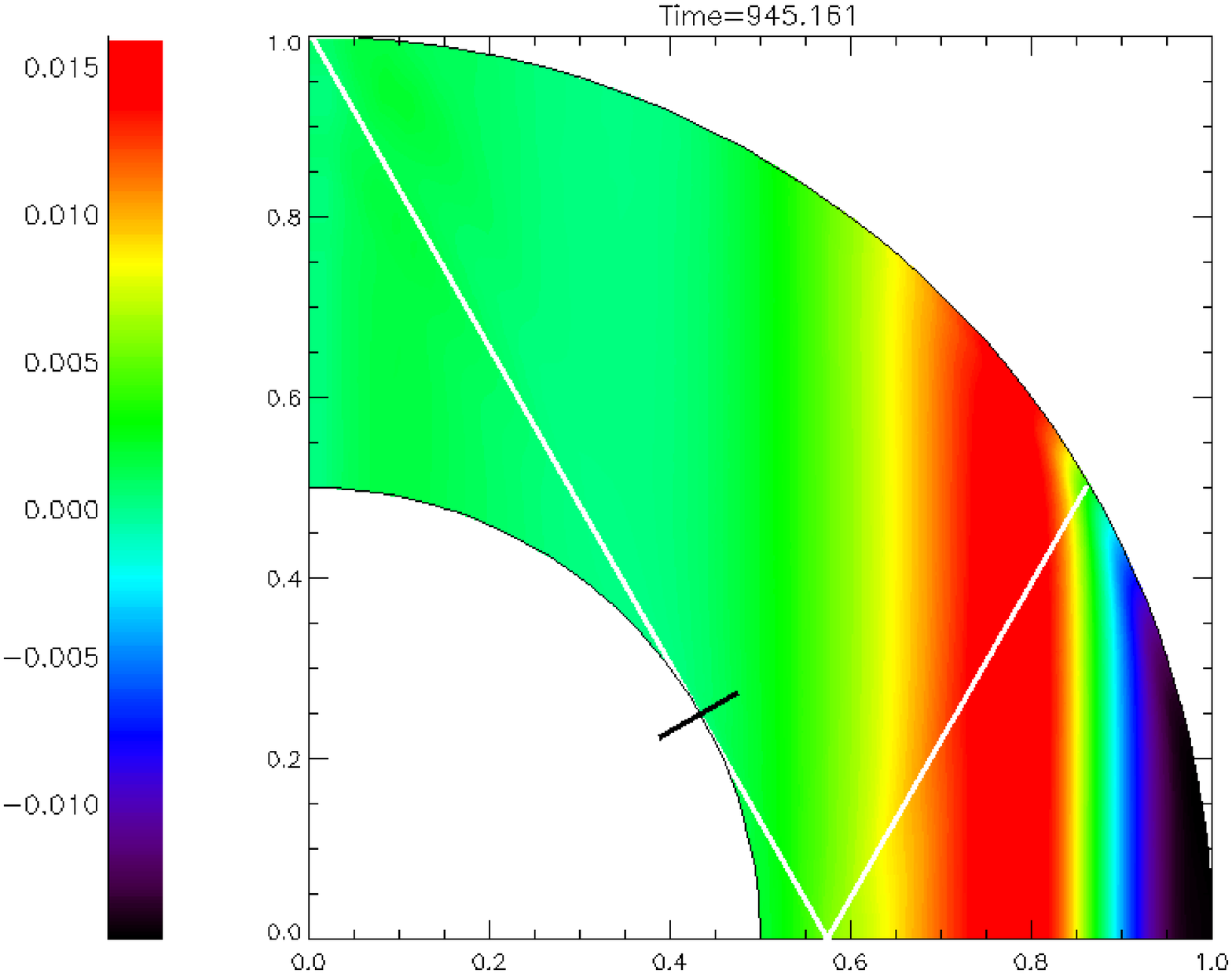}} &
      \resizebox{55mm}{!}{\includegraphics{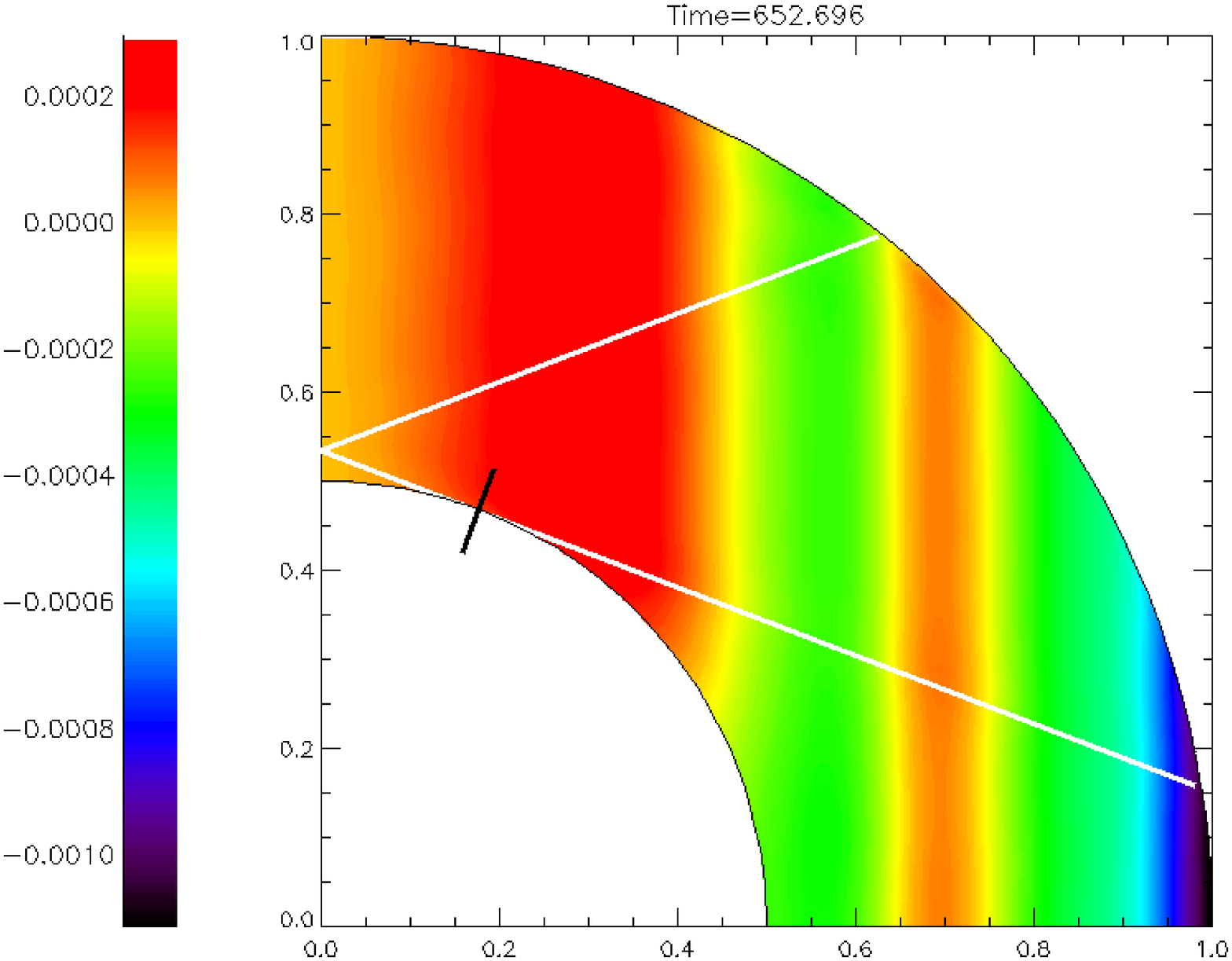}}
    \end{tabular}
    \begin{picture}(10,0)
        \put(-265,120){\Large{$\left<u_{\phi}\right>_{\phi}$}}
        \put(-145,104){$\omega/\Omega=0.6$}
        \put(24,104){$\omega/\Omega=1.0$}
        \put(192,104){$\omega/\Omega=1.87$}
        \put(-145,225){$\omega/\Omega=-0.2$}
        \put(24,225){$\omega/\Omega=-0.8$}
        \put(192,225){$\omega/\Omega=-1.6$}
    \end{picture}
    \caption{Azimuthally averaged azimuthal velocity in a meridional plane at various times for various forcing frequencies.
In all cases, the amplitude of the forcing is $A=10^{-2}$ whereas the Ekman number is $E=10^{-5}$.
Black corresponds to the minimum of the azimuthal velocity whereas red corresponds to the maximum.
The black lines perpendicular to the inner core shows the location of the critical latitude at $t=0$ whereas the white lines shows some of the ray paths emitted from that particular latitude assuming uniform rotation.
This shows that tidal forcing generates cylindrical differential rotation in the body in the synchronisation process, and the fluid certainly does not spin up as a solid body.\label{fig:azimuth}}
  \end{center}
\end{figure*}

As can be seen in Fig.~\ref{fig:ut}, the wave structure observed in Fig.~\ref{fig:u_ex} in the linear regime is still visible in the nonlinear regime, even after thousands of periods. 
In fact, there are barely any differences between the linear and nonlinear regimes when looking at the radial or tangent components of the velocity at the same time.
As will be discussed in Section~\ref{sec:zonal}, the main difference between the linear and nonlinear regimes is related to mean zonal flows.
This is partly because the total rotation rate $\Omega+\delta\Omega$ does not vary significantly during this simulation (see the amplitudes in the bottom panel of Fig.~\ref{fig:amp}).
In addition, the internal shear layers or the flow near the critical latitude do not appear to be unstable to small-scale instabilities.
Instead, the dominant nonlinearities in the regime probed by these simulations, appear to be associated with the generation of zonal flows, as we will now describe.

\subsection{Generation of zonal flows}\label{sec:zonal}

In all of our nonlinear simulations, we observe a non-uniform deposition of angular momentum in the fluid.
Since this angular momentum injection (or extraction) by the forcing is not distributed homogeneously in the fluid, this leads to the generation of differential rotation in the form of zonal flows, which is a generic feature of all nonlinear simulations reported in this paper.
We show in Fig.~\ref{fig:azimuth} an example of such zonal flows for various positive and negative Doppler-shifted frequencies $\omega$.
The azimuthal component of the velocity is azimuthally-averaged and plotted in the meridional plane.
The results shown in Fig.~\ref{fig:azimuth} correspond to $A=10^{-2}$ and $E=10^{-5}$ in all cases, and are plotted at an arbitrary time in the range $500<\Omega t<1300$.
We stress, however, that these results are time-dependent and the azimuthal flows evolve significantly during the course of a given simulation. 

It is clear that the azimuthal flow significantly departs from a purely solid body rotation. 
The white lines in Fig.~\ref{fig:azimuth} correspond to the path of characteristics emitted at the inner critical latitude.
Angular momentum seems to be preferentially deposited where the inertial waves reflect on the outer boundary (this is particularly visible for $\omega/\Omega=-0.2$, $\omega/\Omega=-0.8$ and $\omega/\Omega=1.87$).
This differential rotation affects the local properties of inertial waves, since the frequency $\omega$, and therefore the local direction of propagation of inertial waves, now depends on the cylindrical radius.
The azimuthal component of the velocity is indeed nearly vertically invariant in most of our simulations, corresponding to geostrophic flows with rotation constant on cylinders.
The properties of small-amplitude inertial waves propagating in a differentially rotating incompressible fluid contained in a spherical shell have been recently investigated by \cite{baruteau2013}, who note that inertial waves can now propagate along curved paths since the Doppler-shifted frequency is now a function of space.
\begin{figure}
  \begin{center}
    \begin{tabular}{c}
\resizebox{80mm}{!}{\includegraphics{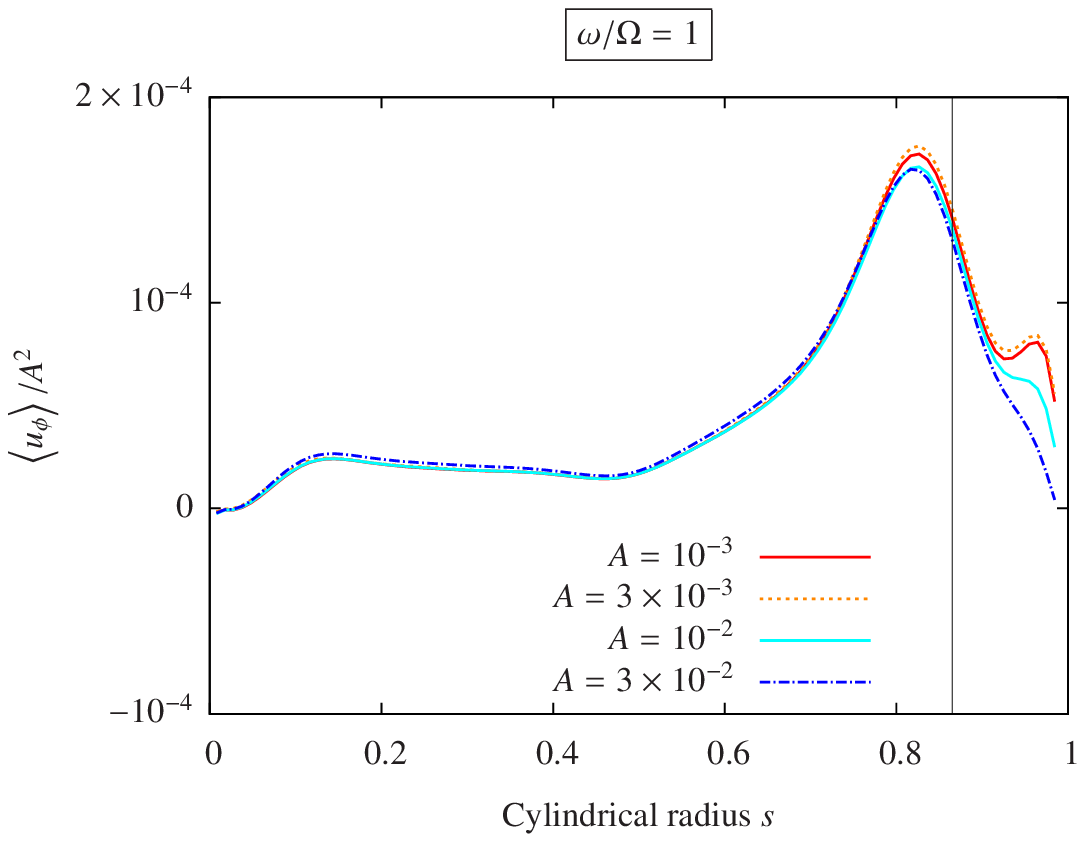}}\\
\resizebox{80mm}{!}{\includegraphics{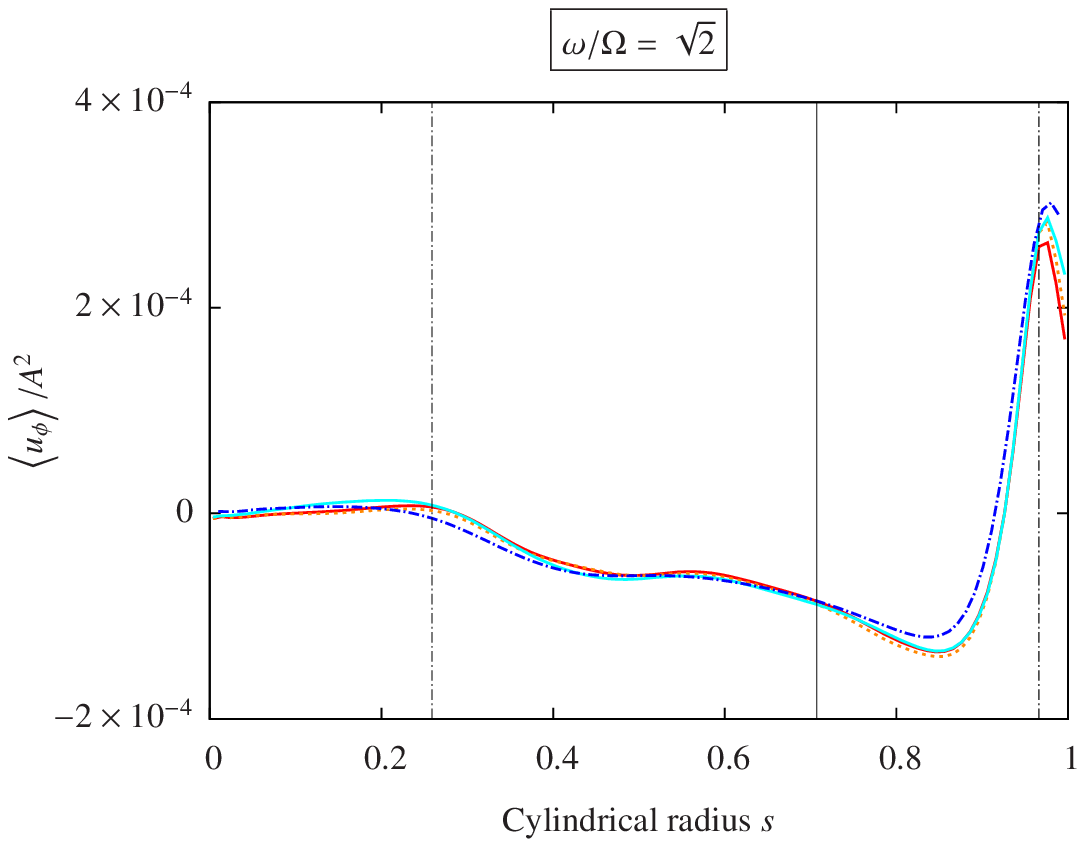}}
    \end{tabular}
    \caption{Averaged zonal velocity for $\omega/\Omega=1$ and $\omega/\Omega=\sqrt{2}$ and various amplitudes of the forcing. $\left<.\right>$ corresponds to a vertical and azimuthal average. Both plots correspond to $\Omega t=50$. The zonal velocities are scaled with the square of the amplitude of the boundary forcing. The two vertical dotted lines correspond to the location of the reflections (see Fig.~\ref{fig:u_ex}) on the outer boundary whereas the solid line corresponds to the critical latitude on the outer boundary.\label{fig:scaling_amp}}
  \end{center}
\end{figure}

It has been suggested that the interactions of $m=2$ inertial modes in a spherical shell can generate a zonal flow \citep{tilgner2007}.
As expected from the weakly nonlinear origin of these zonal flows, their amplitudes scale as the square of the amplitude of the forcing.
This is confirmed by the results presented in Fig.~\ref{fig:scaling_amp}.
The azimuthal component of the velocity is averaged along the vertical and azimuthal directions and plotted against the cylindrical radius $s=r\sin\theta$.
We plot the results for two different frequencies, $\omega/\Omega=1$ and $\sqrt{2}$, at $E=10^{-5}$ and various amplitudes.
All results corresponds to the early time $\Omega t = 50$.
As the zonal velocities are scaled with $A^2$, they all collapse onto approximately the same curve.
Note that this scaling was already observed in Fig.~\ref{fig:amp} at early times.
At later times, the complicated frequency dependence of the inertial wave excitation, as observed in the linear problem, makes these solutions depart from each other more strongly.
\begin{figure}
  \begin{center}
    \begin{tabular}{c}
      \resizebox{80mm}{!}{\includegraphics{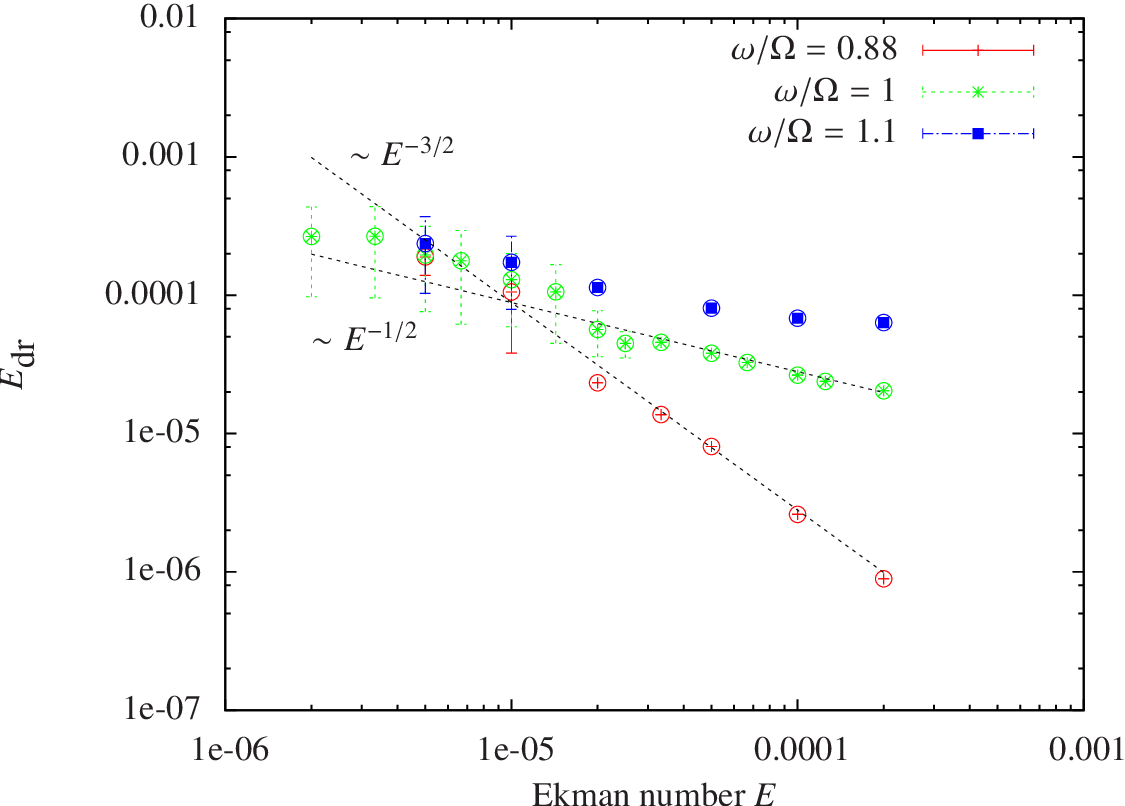}}
    \end{tabular}
    \caption{Energy in the differential rotation as defined by equation \eqref{eq:edr}. The error bars correspond to the maximum and minimum values between $\Omega t=10^3$ and $\Omega t=10^4$. A large Ekman numbers (\textit{i.e.} $E>10^5$), a nonlinear steady state is reached for $\Omega t > 10^3$ which is why no error bars are present in that case.\label{fig:edr}}
  \end{center}
\end{figure}

These results also give an indication as to where the nonlinearities are dominant.
The case $\omega/\Omega=1$ is peculiar, since the location where the waves reflect on the outer boundary is the critical latitude of the outer sphere, \textit{i.e.} $s=0.87$.
In any case, the angular momentum seems to be dominantly deposited near to that radius.
The case $\omega/\Omega=\sqrt{2}$ is more representative, since the locations of the wave reflection (see Fig.~\ref{fig:u_ex}) are distinct from the outer critical latitude.
The former are shown in Fig.~\ref{fig:scaling_amp} as dotted lines whereas the latter is shown as a solid line.
Clearly, the wave reflections on the outer boundary close to the equator play a dominant role compared to the outer critical latitude.

The mechanism responsible for the zonal flows in our case is at odds with the no-slip librating case where it has been shown that the torque in the Ekman boundary layers and its eruption at the outer critical latitude are dominant and drive the zonal flow \citep{calkins2010}.
In that case, the amplitude of the zonal flow scales as the square of the amplitude of the forcing but is independent of the Ekman number.
In our case however, there are no Ekman boundary layers since we adopt stress-free boundaries.
We therefore do no expect the same mechanism to drive zonal winds.
As already mentioned, the reflection of the waves generated at the inner critical latitude on the outer boundary seems to be responsible for the generation of the zonal flow.
While our result depends strongly on the nature of wave reflection at the boundaries, it is known that these reflections are very similar between a solid wall and a free surface as long as the frequency is much smaller than the one of surface gravity waves \citep{phillips1963}.
However, since we allow for a radial flow across the outer boundary, it is not clear to what extent our choice of boundary conditions influences the nonlinear behaviour of the waves.
This can only be resolved by considering a more realistic ellipsoidal geometry or by changing the way the waves are forced.
This is left for future work.

We now attempt to quantify the variation of the amplitude of the differential rotation as a function of the Ekman number.
It is difficult to define the amplitude of the differential rotation at low Ekman numbers, since there is generally no quasi-steady state.
Nevertheless, it is still helpful to define the energy in the differential rotation as \citep{tilgner2007},
\begin{equation}
\label{eq:edr}
E_{\textrm{dr}}=\frac12\int_V\left[\left<u_{\phi}\right>_{\phi}-\delta\Omega \ \! r\sin\theta\right]^2\textrm{d}V \ ,
\end{equation}
where $\left<u_{\phi}\right>_{\phi}$ is the azimuthal average of the zonal velocity and $\delta\Omega$ is the rotation rate of the fluid at a particular time, as seen in the frame rotating at the rate $\Omega$.
In Fig.~\ref{fig:edr}, we plot this quantity versus the Ekman number for $A=10^{-2}$ and three different frequencies $\omega/\Omega=0.88$, $\omega/\Omega=1$ and $\omega/\Omega=1.1$.
At large enough Ekman numbers (\textit{i.e.} $E>5\times10^{-4}$), the system can reach a nonlinear steady state allowing for an unambiguous measure of the amplitude of the zonal flows.
As one decreases the Ekman number, the system becomes periodic or chaotic which explains the presence of error bars in Fig.~\ref{fig:edr}.
In any case, the energy in the differential rotation increases as a negative power of the Ekman number, as is also found in the numerical model of \cite{tilgner2007} and in the experiments of a coreless deformed sphere by \cite{sauret} and \cite{sauret2013b}.
The two straight lines in Fig.~\ref{fig:edr} correspond to the arbitrary scalings $E^{-3/2}$ and $E^{-1/2}$, and are shown for illustration.
\textcolor{black}{Note that the amplitude of forcing of $A=0.01$ that we are considering in Fig.~\ref{fig:edr} can be considered to roughly represent the amplitude of the synchronisation tide inside a hot Jupiter orbiting a solar-type star in a one day orbit (if the tidal frequency is comparable with the spin frequency). A naive extrapolation of our results (for any scaling exponent in the given range) inside a hot Jupiter (taking an approximate value of $E\approx10^{-18}$) would suggest that the energy in the differential rotation would far exceed that associated with the solid body rotational kinetic energy, i.e. very large shears would be predicted. However, it is very likely that shear instabilities would become important as the Ekman number is decreased (see section \ref{sec:shear} below), thereby modifying the scaling behaviours and preventing such large shears from developing. In addition, angular momentum redistribution by turbulent convection or magnetic fields, which we have omitted, could limit the amplitude of these zonal flows.}

\begin{figure}
  \begin{center}
    \begin{tabular}{c}
      \resizebox{90mm}{!}{\includegraphics{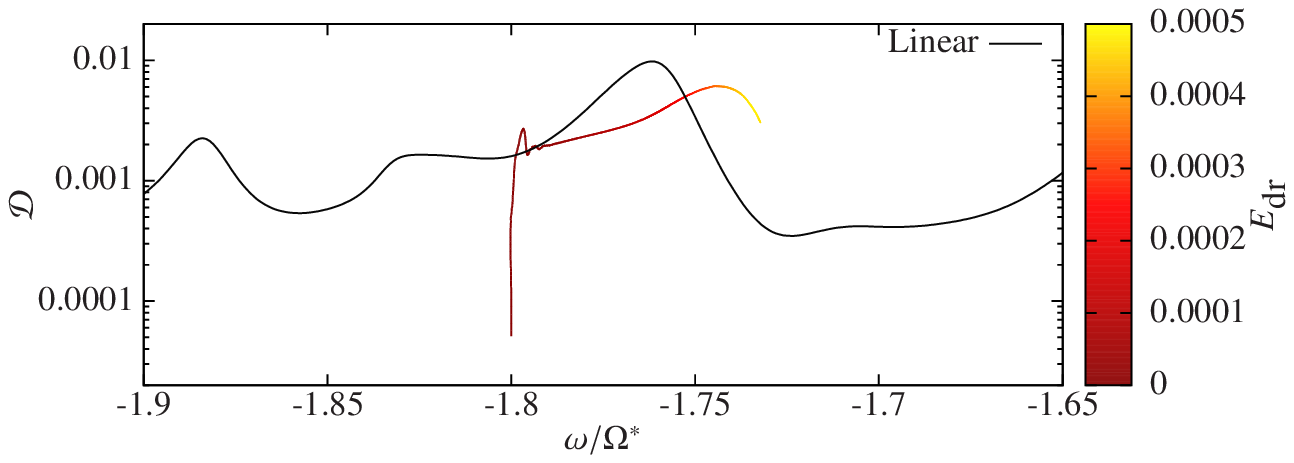}} \\
      \resizebox{90mm}{!}{\includegraphics{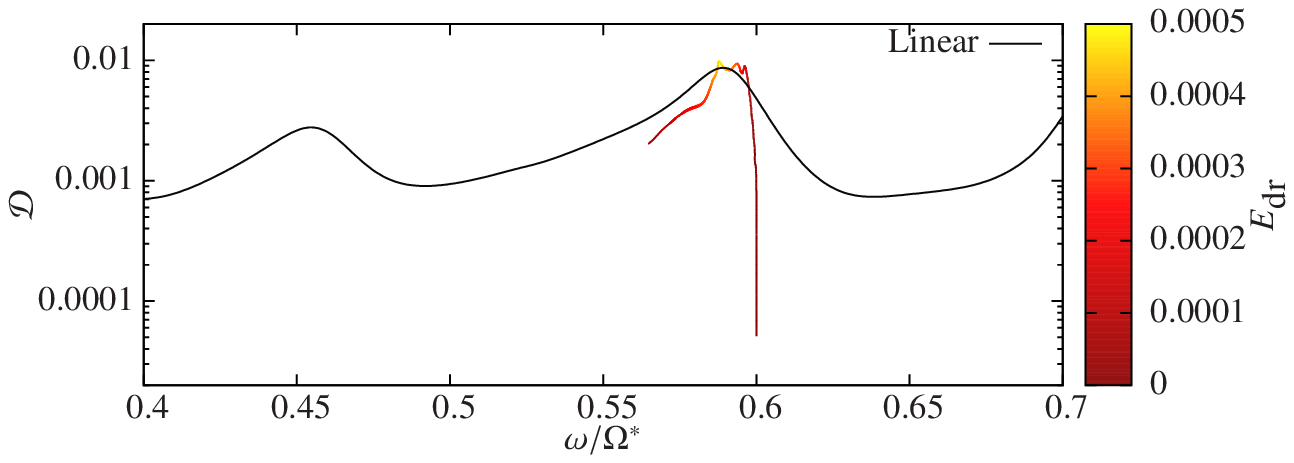}} \\
      \resizebox{90mm}{!}{\includegraphics{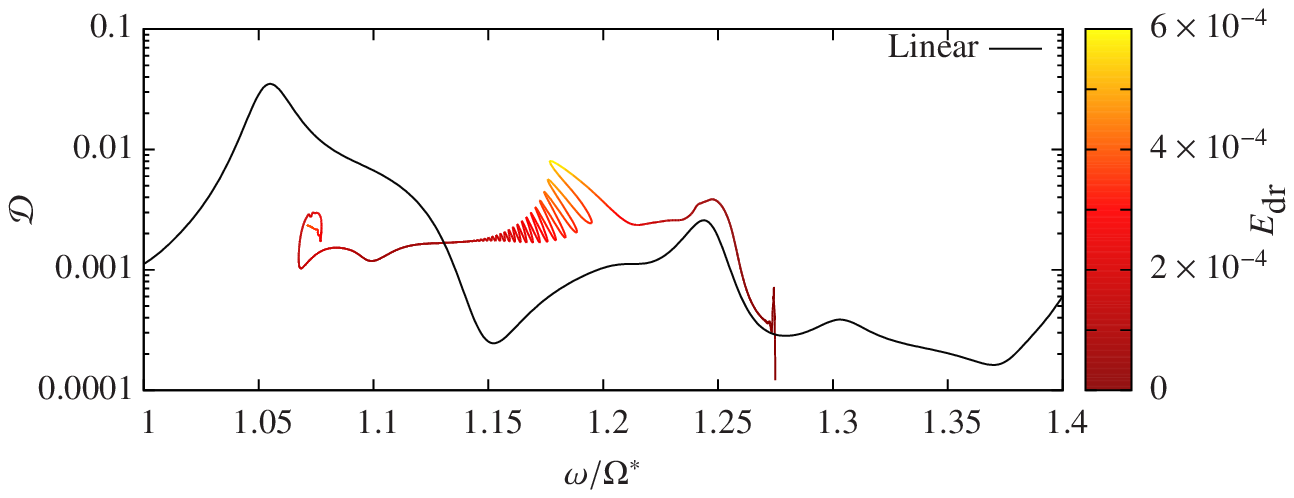}}
    \end{tabular}
    \caption{Dissipation rate as a function of $\omega/\Omega^*$, where $\Omega^*=\Omega+\delta\Omega$ is the total rotation rate of the fluid. The linear prediction is shown as a black line whereas the time evolution obtained from nonlinear simulations is shown as a colored line, where the color corresponds to the amplitude of the differential rotation, as defined by equation \eqref{eq:edr}. From top to bottom, the initial Doppler-shifted frequencies are $\omega/\Omega=-1.8$, $\omega/\Omega=0.6$ and $\omega/\Omega=1.275$\label{fig:sp}}
  \end{center}
\end{figure}

\textcolor{black}{Finally, in order to show that the differential rotation is indeed the dominant nonlinear feature observed in our simulations, we further compare our nonlinear results with predictions obtained from direct linear calculations. In particular, we focus on the total viscous dissipation rate calculated as a function of the Doppler-shifted frequency of the forcing. During the course of our nonlinear simulations, we monitor the total rotation rate of the fluid $\Omega^*=\Omega+\delta\Omega$, the viscous dissipation rate and the energy in the differential rotation as defined by equation \eqref{eq:edr}. In Fig.~\ref{fig:sp}, we plot the dissipation rate as a function of the normalised frequency $\omega/\Omega^*$. In the linear case, the rotation rate of the fluid is fixed and the curved is obtained by direct linear calculation for each value of $\omega/\Omega^*$ (using the same method as in section \ref{sec:comp_gordon} and in \citealt{ogilvie2009}). In the nonlinear case, the total rotation rate and the viscous dissipation rate are changing with time. We show the results corresponding to three different initial frequencies: $\omega/\Omega=-1.8$, $\omega/\Omega=0.6$ and $\omega/\Omega=1.275$. In all cases, we observe an initial transient phase where the dissipation rapidly increases (as observed in Fig.~\ref{fig:amp} and Fig.~\ref{fig:freq}). The dissipation then saturates close to the value predicted by linear theory. As time increases, the differential rotation builds up and the dissipation rate predicted by our nonlinear simulations departs from that predicted by linear theory at the same Doppler-shifted frequency. This indicates that the linear predictions are accurate in the early stage of the synchronisation process, but that the nonlinear path to synchronisation will be very different, for example due to the differential rotation driven by nonlinearities. Note however that for some frequencies, departure or similarities between linear and nonlinear predictions are observed irrespective of the differential rotation (see $\omega/\Omega=0.6$ in Fig.\ref{fig:sp} for example, where a relatively strong differential rotation does not result in a dissipation rate very different from linear predictions).}

\subsection{Varying the frequency}

\begin{figure}
  \begin{center}
    \begin{tabular}{c}
      \resizebox{80mm}{!}{\includegraphics{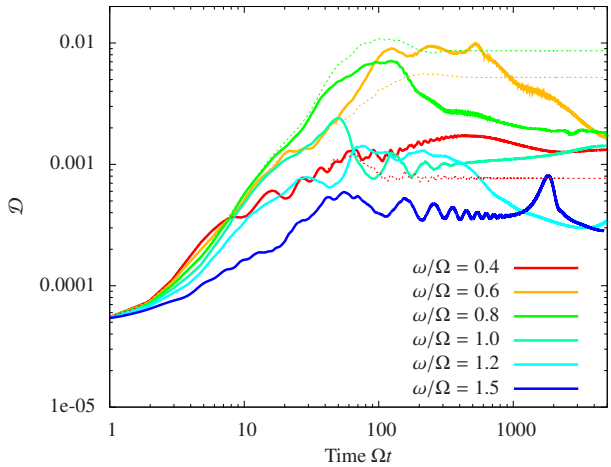}}
    \end{tabular}
    \caption{Dissipation rate (in units of $\rho r_e^3 A^2 \Omega$) versus time for various positive frequencies, $E=10^{-5}$ and $A=10^{-2}$. For $\omega/\Omega=0.4$, $0.6$ and $0.8$, we also plot the linear results as dotted lines for comparison.\label{fig:freq}}
  \end{center}
\end{figure}

While it is not possible to consider every frequency within the range $-2<\omega/\Omega<2$ in the nonlinear case, we consider several frequencies while fixing the forcing amplitude to be $A=10^{-2}$.
The viscous dissipation rates versus time are shown in Fig.~\ref{fig:freq}.
Note that for all frequencies considered here, no steady state is obtained for $E=10^{-5}$ even after $5000$ periods.
For $\omega/\Omega=0.4$, $0.6$ and $0.8$, we also plot the linear results as dotted lines for comparison.
The dissipation in the nonlinear regime can be larger than in the linear regime (as it is the case for $\omega/\Omega=0.4$), smaller (as it is the case for $\omega/\Omega=0.8$) or both depending on time.
The complicated dependence of the viscous dissipation on frequency was already observed in the linear regime \citep{ogilvie2009} and is also a property of the nonlinear regime.
This is related to the fact that the dynamics is still strongly dominated by wave beams generated at the inner critical latitude and reflecting on the boundaries.
\begin{figure*}
  \begin{center}
    \begin{tabular}{ccc}
      \resizebox{60mm}{!}{\includegraphics{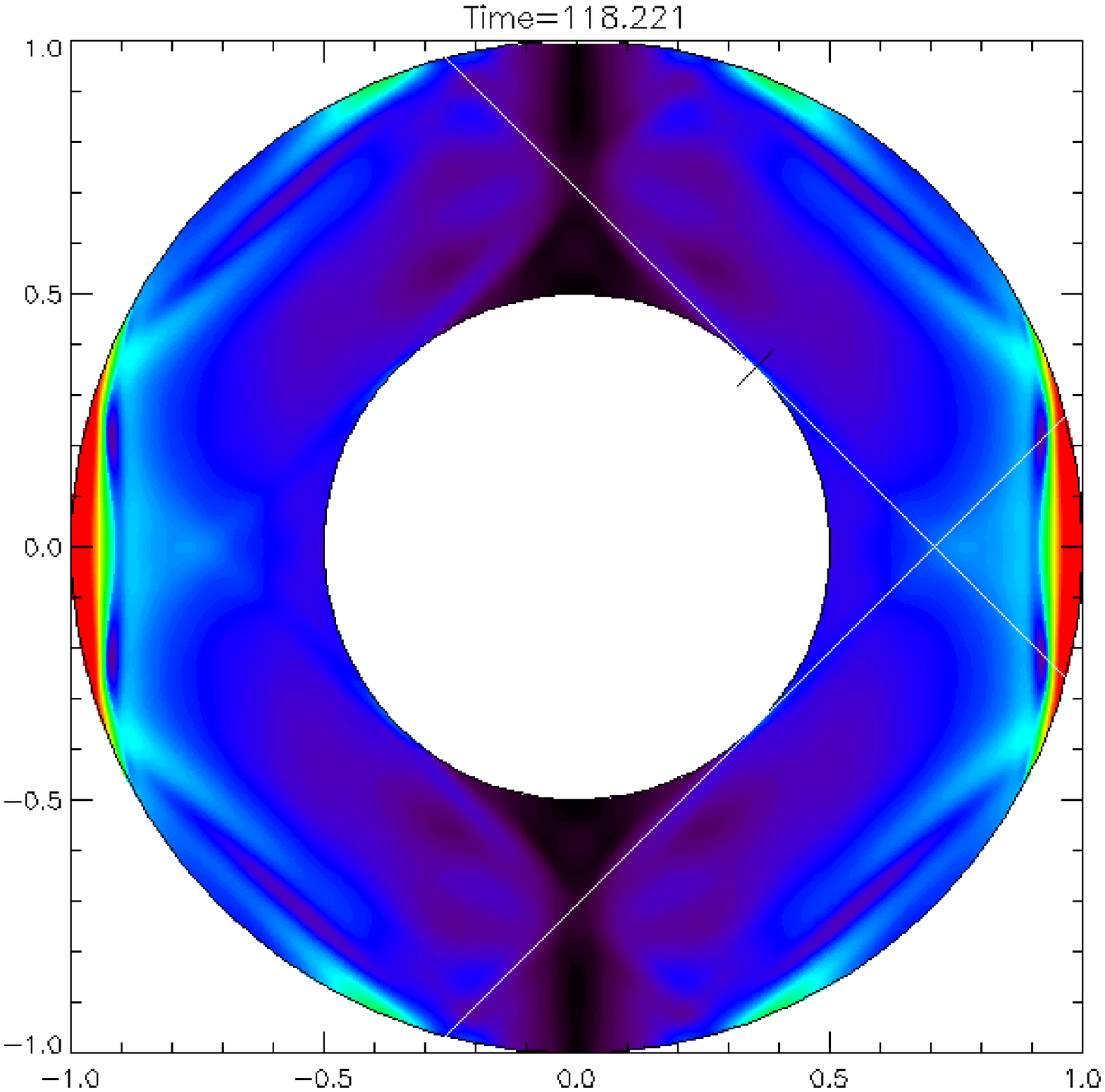}}
      \resizebox{60mm}{!}{\includegraphics{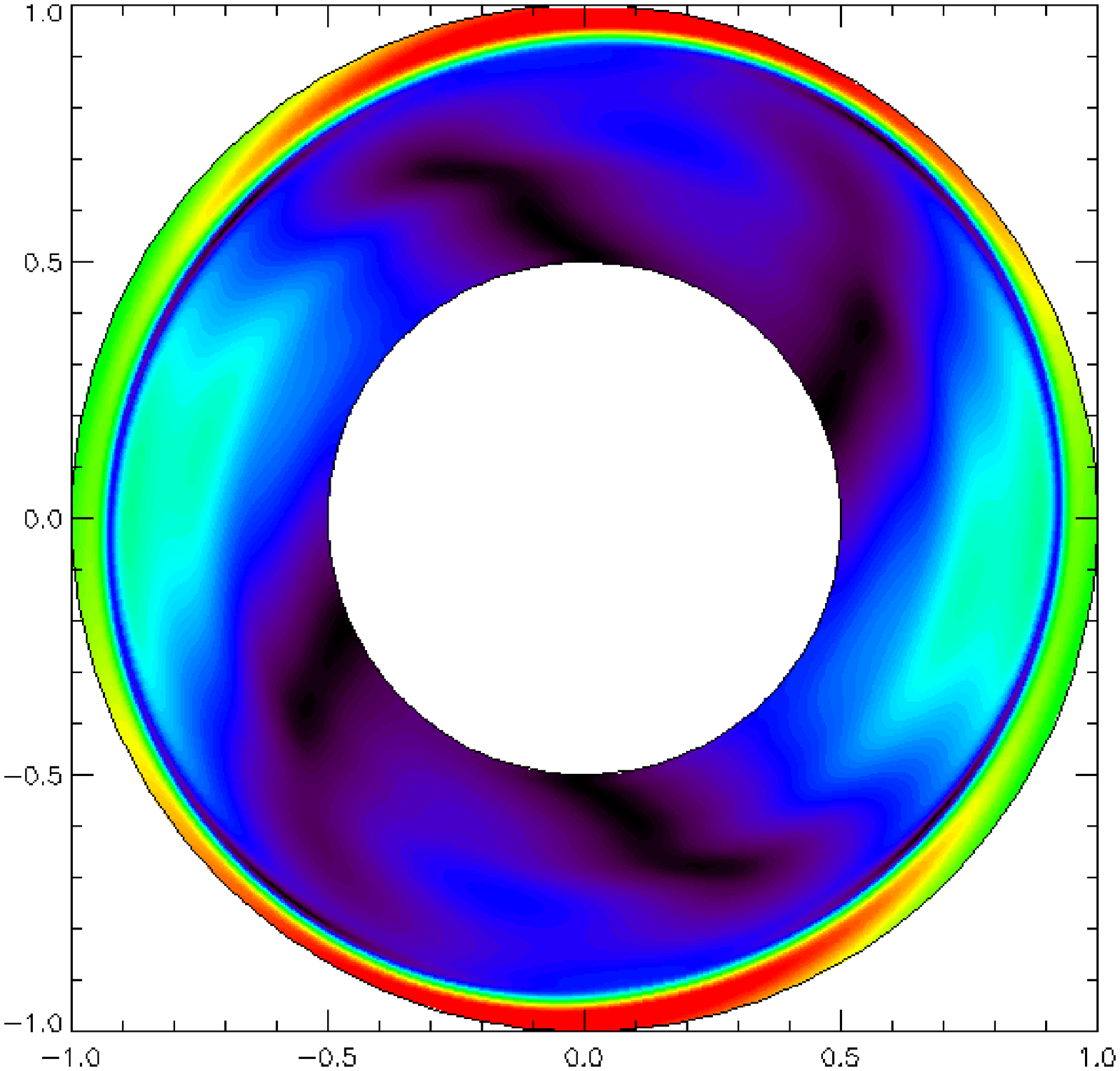}}
      \resizebox{48mm}{!}{\includegraphics{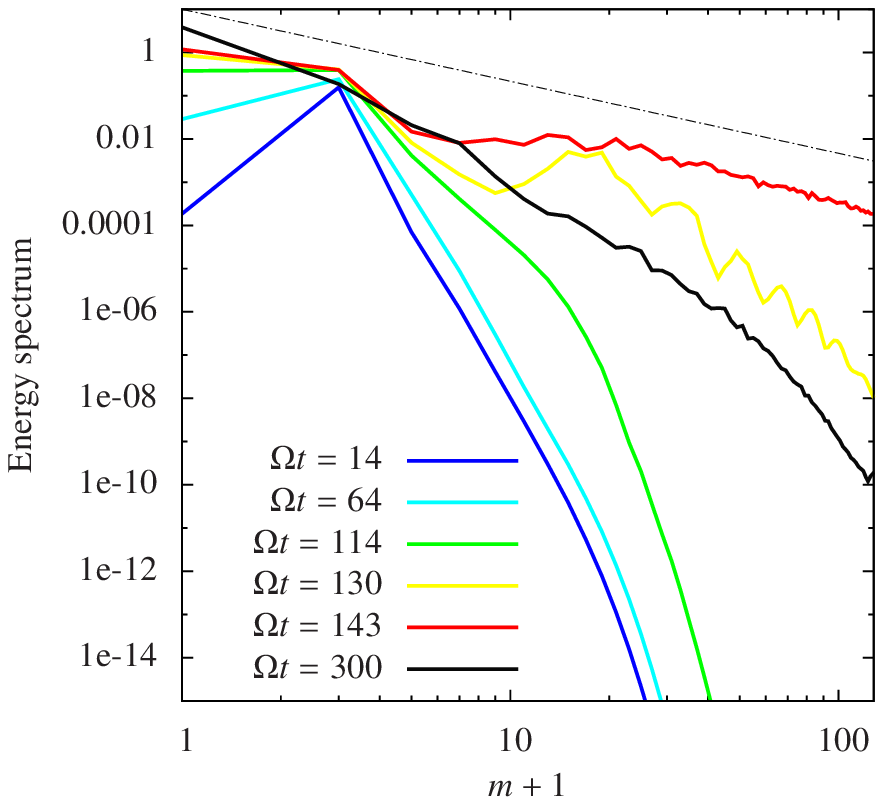}} \\
      \resizebox{60mm}{!}{\includegraphics{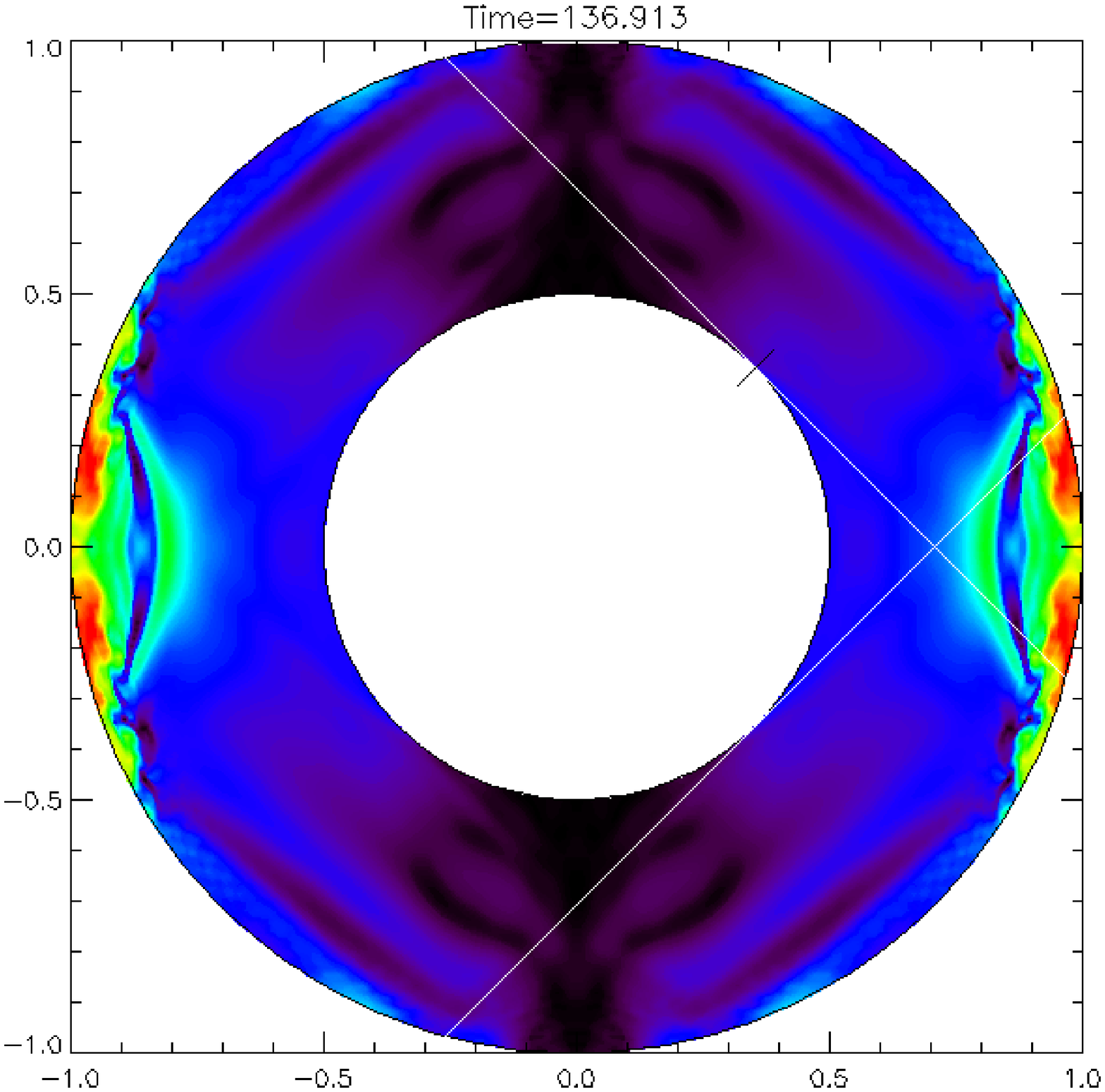}}
      \resizebox{60mm}{!}{\includegraphics{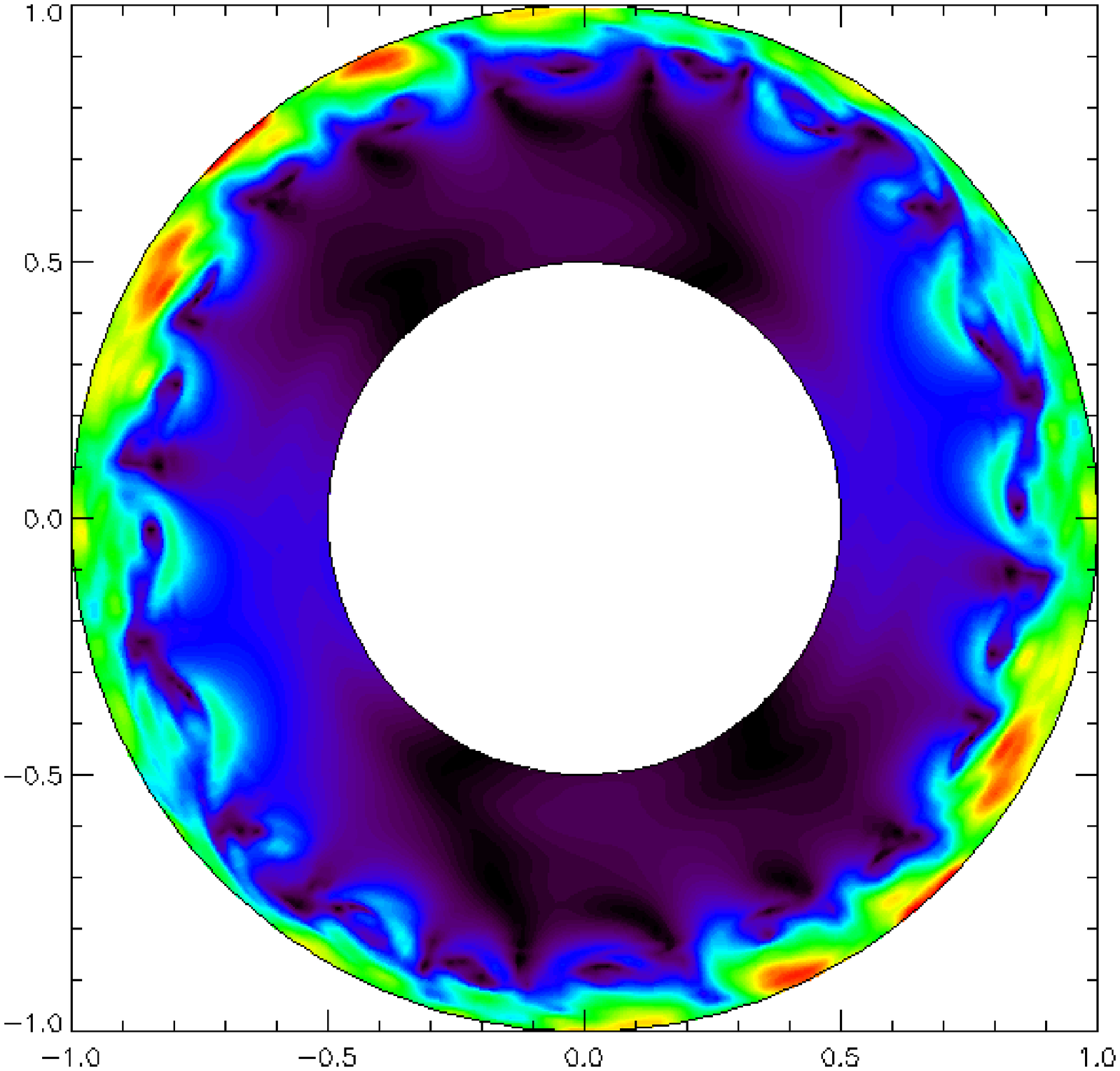}}
      \resizebox{50mm}{!}{\includegraphics{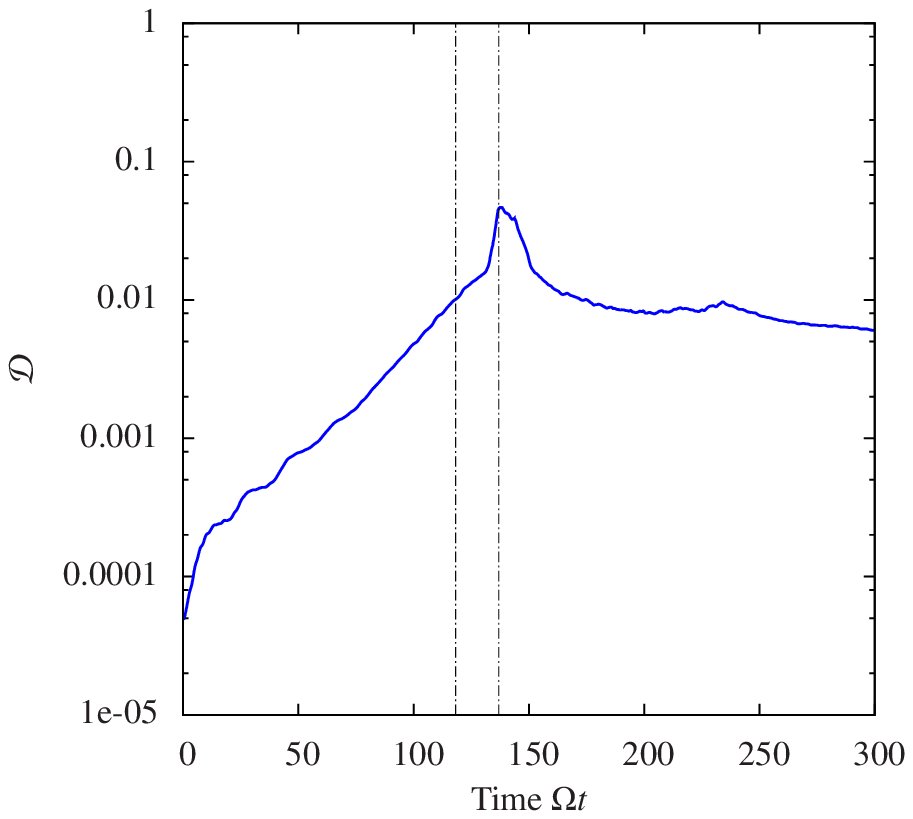}}
    \end{tabular}
    \caption{Secondary shear instability. We show the velocity amplitude in meridional and equatorial planes just before ($\Omega t\approx118$) and after ($\Omega t\approx 137$) the instability. The kinetic energy spectrum as a function of the shifted azimuthal wave number $m+1$ is shown on the top right. The thin dotted line corresponds to the slope $m^{-5/3}$. The time evolution of the dissipation rate is shown on the bottom right. The two vertical dotted lines correspond to the two times at which the flow is visualised on the left.\label{fig:shear}}
  \end{center}
\end{figure*}

\subsection{Secondary shear instabilities\label{sec:shear}}

We have observed the amplitude of the zonal flows to increase as we decrease the Ekman number in Fig.~\ref{fig:edr}.
This suggests that strong localised shear could be generated in the regime of very small Ekman numbers, which might be unstable to hydrodynamical shear instabilities.
In this section we present the results from a particular simulation with $A=0.02$, $\omega/\Omega=\sqrt{2}$ and $E=10^{-5}$.
As already observed, we find an increase in the vertical component of the angular momentum, and a zonal flow is driven, primarily close to the location of the reflection on the outer boundary, which in this case in close to the equator.
We show on the left four panels of Fig.~\ref{fig:shear} the amplitude of the velocity $|\bm{u}|$ is a meridional slice and in the equatorial plane at $\Omega t \approx 118$ and $\Omega t \approx 137$.
We also show the time evolution of the total viscous dissipation on the bottom right panel of Fig.~\ref{fig:shear}.
At time $\Omega t \approx130$, we observe a sudden increase in the total dissipation.
This corresponds to the onset of a hydrodynamical shear instability, which preferentially excites modes with large azimuthal wave numbers $m$.
The kinetic energy spectrum in the azimuthal direction is shown on the top right panel of Fig.~\ref{fig:shear}.
We observe the growth of large $m$ components in the flow in both the spectrum and the equatorial slice.
Large $m$ components were not present during the early nonlinear evolution, where the slope is much steeper in spectral space, indicating only weak nonlinear transfers to modes with smaller azimuthal scales than $m=2$.
The axisymmetric $m=0$ component of the flow continuously grows until energy is transferred into these non-axisymmetric modes, after which is appears to saturate.
After the instability saturates, the energy at small azimuthal wavelength is dissipated, leading to a steeper kinetic energy spectrum (see the spectrum at $\Omega t \approx 300$ in Fig.~\ref{fig:shear}).
Although we didn't pursue the simulation further, it is possible that once the turbulence is dissipated, another shear instability kicks in, leading to a cyclic behaviour between the laminar and turbulent states.

To our knowledge, this is the first numerical evidence of unstable zonal flows driven by inertial waves in spherical shells.
Similar secondary shear instabilities have been recently observed experimentally in a rotating deformed sphere with or without an inner core \citep{sauret,sauret2013b}.
In an astrophysical object, it is possible that magnetic stresses, magnetohydrodynamical shear instabilities, or convection could suppress this differential rotation before it could become unstable to such a hydrodynamical shear instability.

\subsection{No-slip inner core}\label{sec:noslip}
 
\begin{figure}
    \begin{tabular}{c}
      \hspace{-2mm}\resizebox{75mm}{!}{\includegraphics{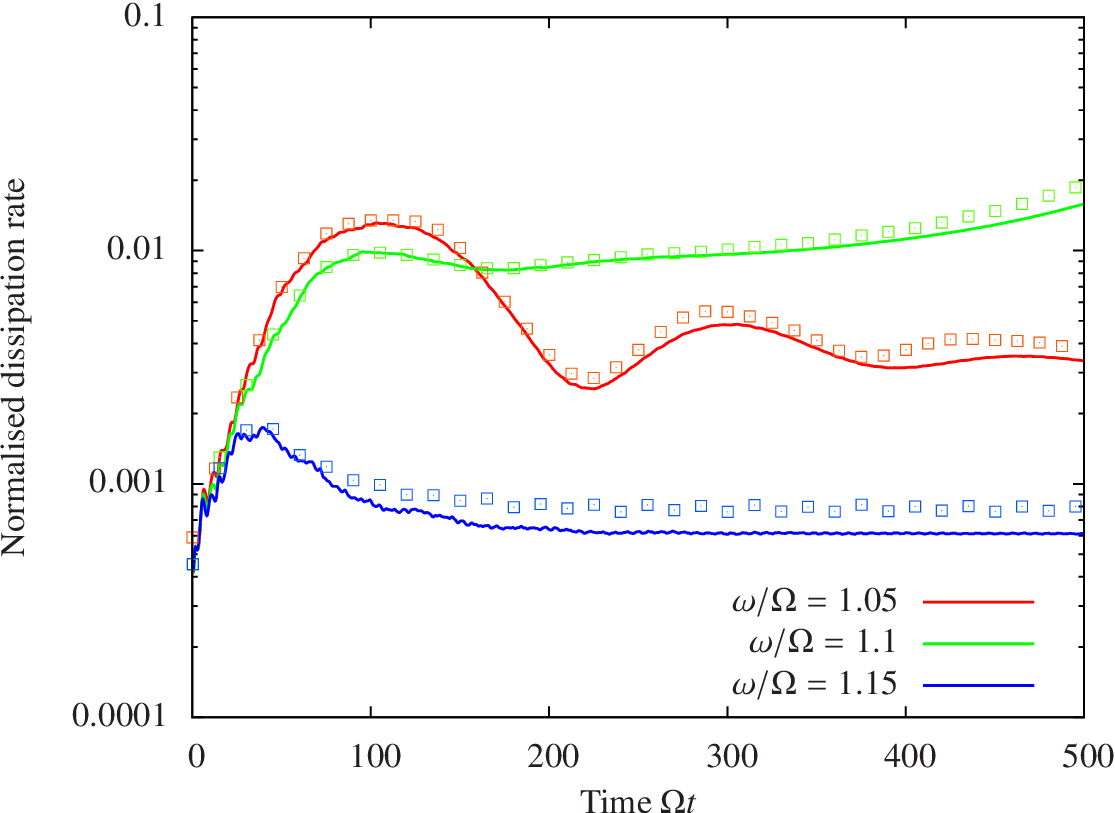}} \\
      \hspace{3mm}\resizebox{70mm}{!}{\includegraphics{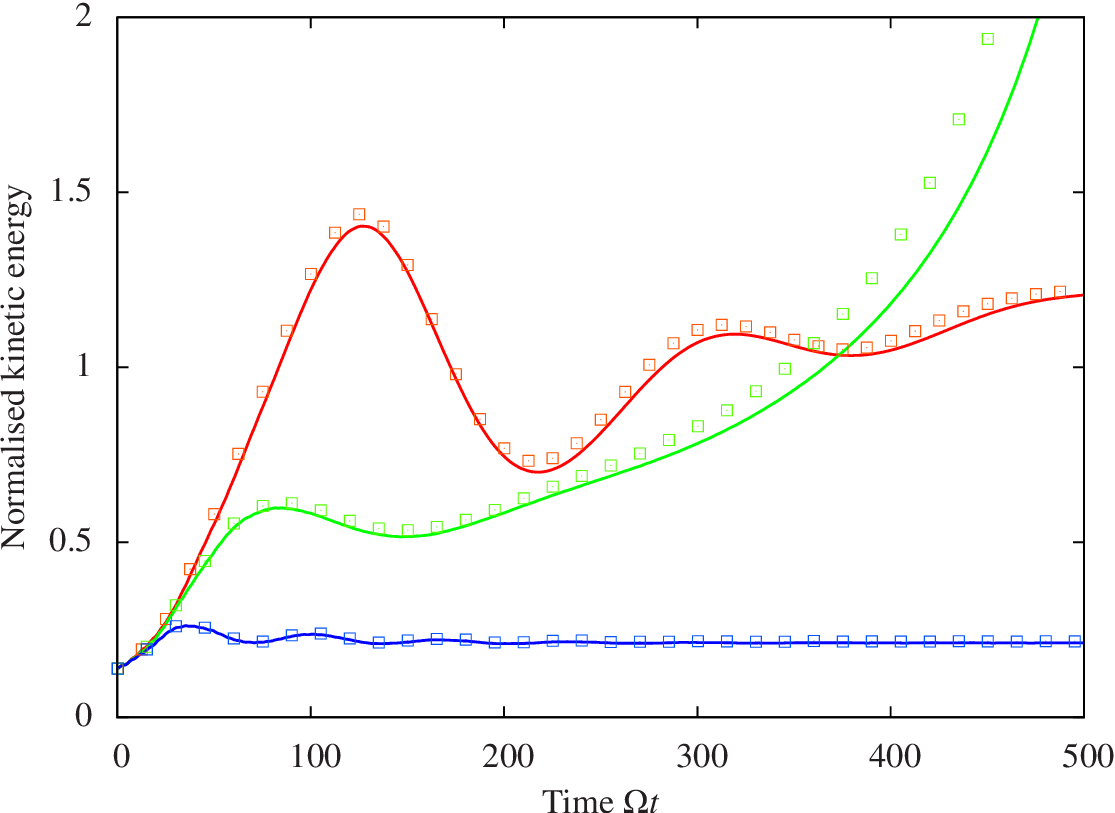}}
    \end{tabular}
  \caption{Same as in Fig.~\ref{fig:3freqcomp} but with no-slip boundary condition at the inner core.\label{fig:noslip}}
\end{figure}

In this section we briefly present our results varying the inner boundary condition on the core.
In particular, we choose to adopt no-slip, rather than stress-free conditions. 
This might be more relevant to terrestrial planets and the solid cores of a giant planet, as well as to laboratory experiments.
In Fig.~\ref{fig:noslip}, we plot the normalised kinetic energy $K$ and viscous dissipation rate $\mathcal{D}$ for a set of nonlinear simulations with $A=3.86\times10^{-3}$, for the three frequencies $\omega/\Omega=1.05$, $1.10$ and $1.15$.
Again, we compare the results from both numerical methods described in Section~\ref{sec:num}.
The kinetic energy evolves very similarly to the stress-free core case.
There are some differences in $\mathcal{D}$, primarily due to the presence of oscillatory Ekman boundary layers that cause periodic oscillations in $L_{x}$ and $L_{y}$.
The similarity between Fig.~\ref{fig:3freqcomp} and Fig.~\ref{fig:noslip} however indicates that the generic nonlinear properties that we have observed in the rest of this paper seem to be robust to this change in the inner boundary condition.

\begin{figure}
    \begin{tabular}{cc}
      \hspace{-3mm}\resizebox{45mm}{!}{\includegraphics{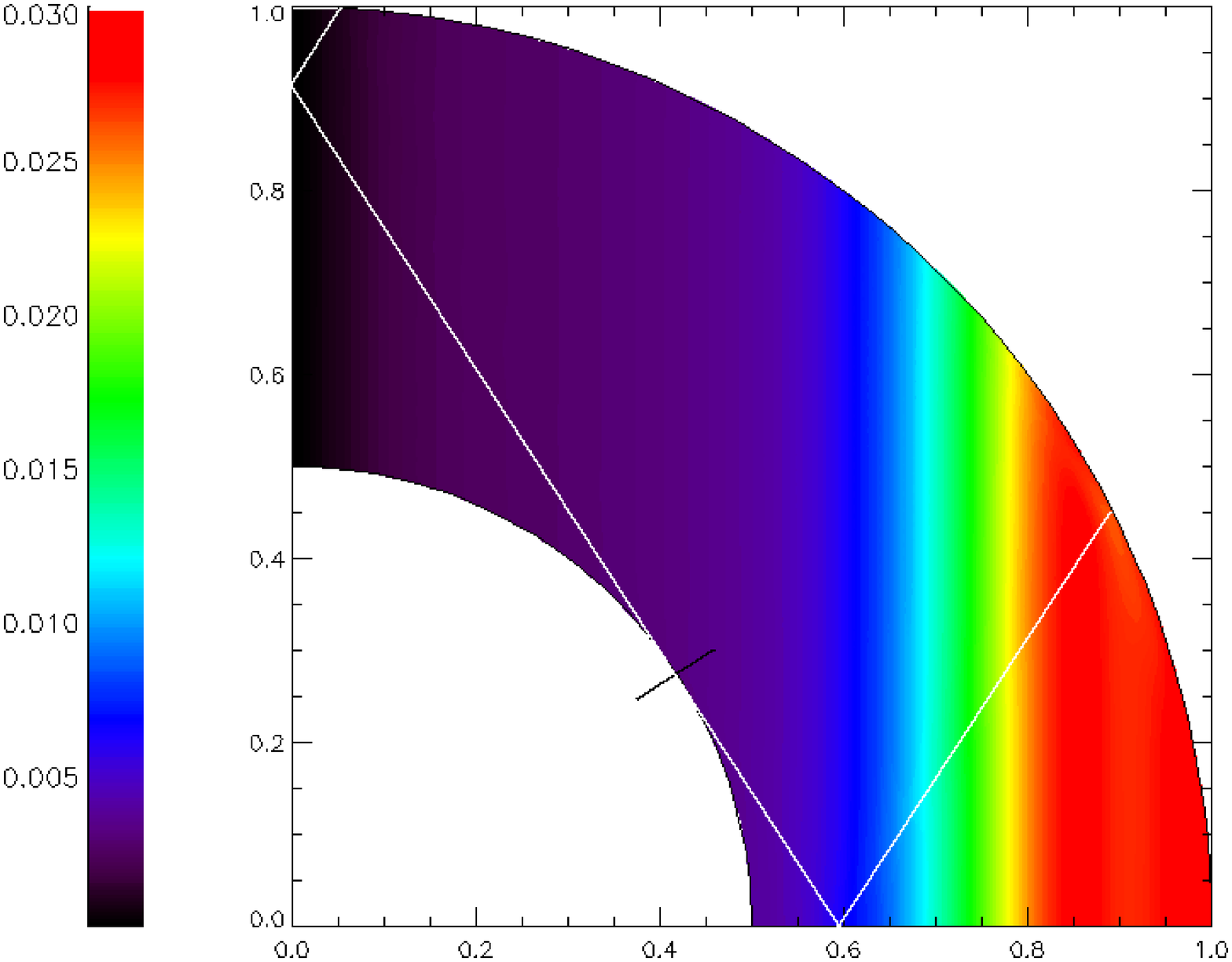}}
      \hspace{-3mm}\resizebox{45mm}{!}{\includegraphics{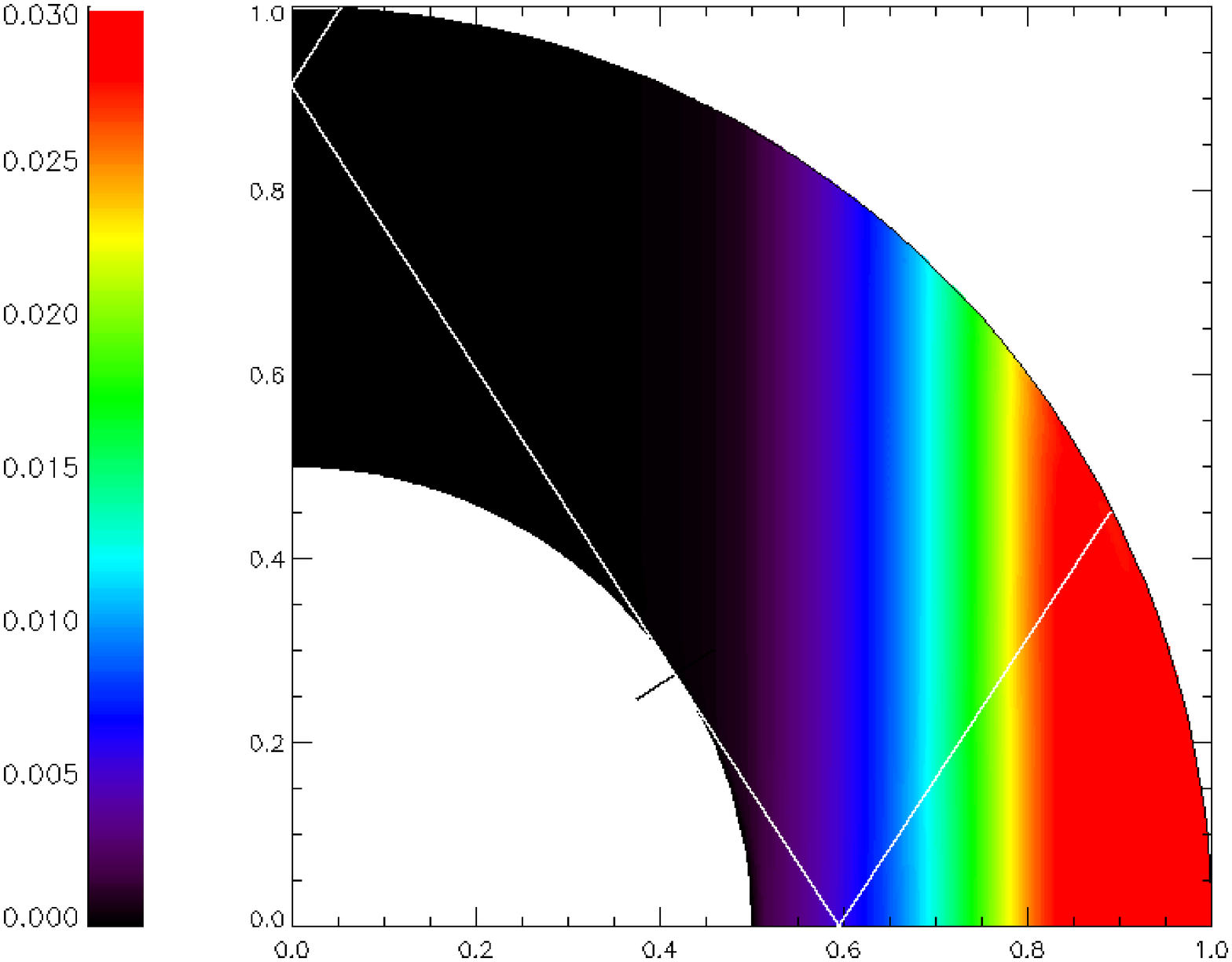}} \\
      \hspace{-3mm}\resizebox{45mm}{!}{\includegraphics{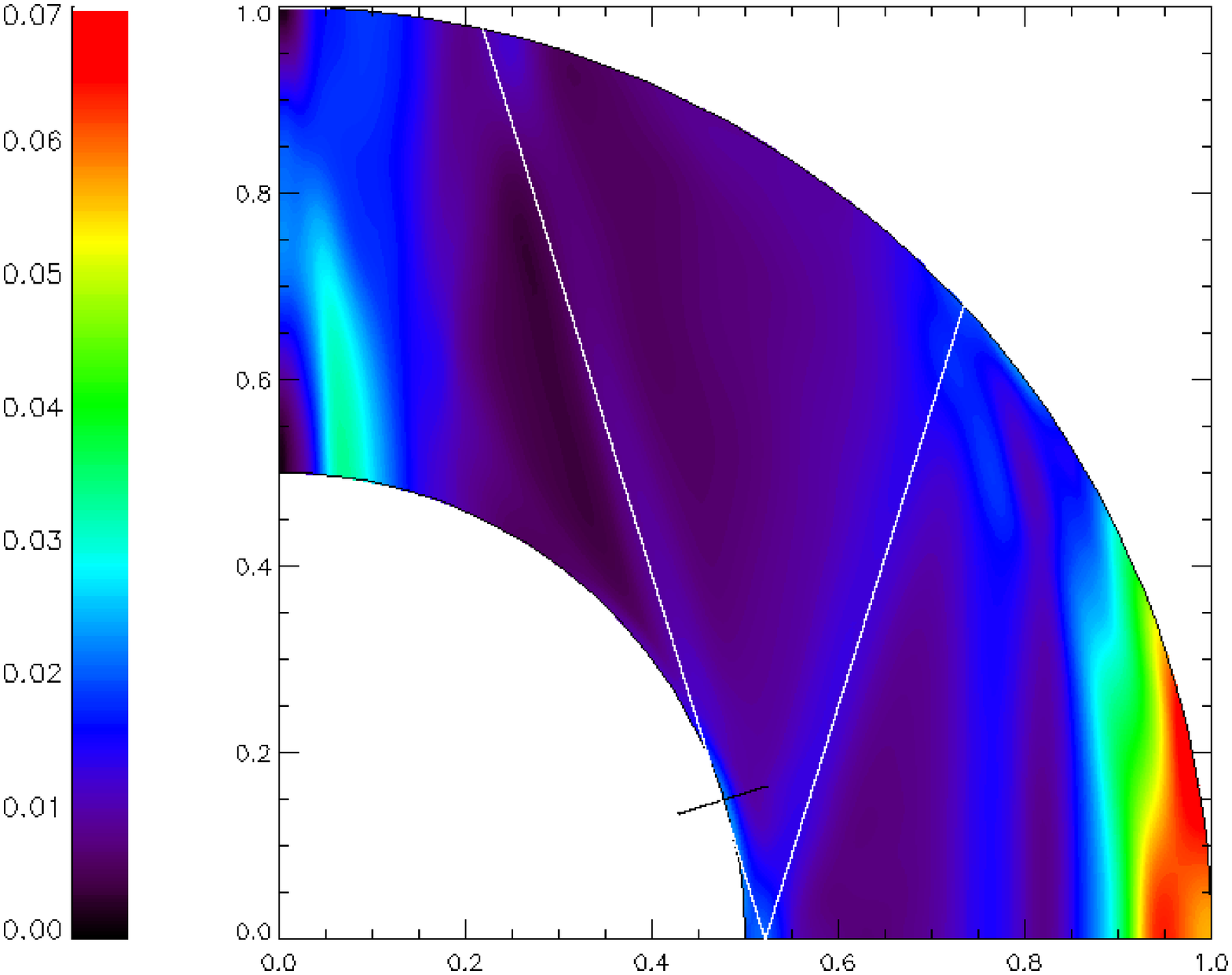}}
      \hspace{-3mm}\resizebox{45mm}{!}{\includegraphics{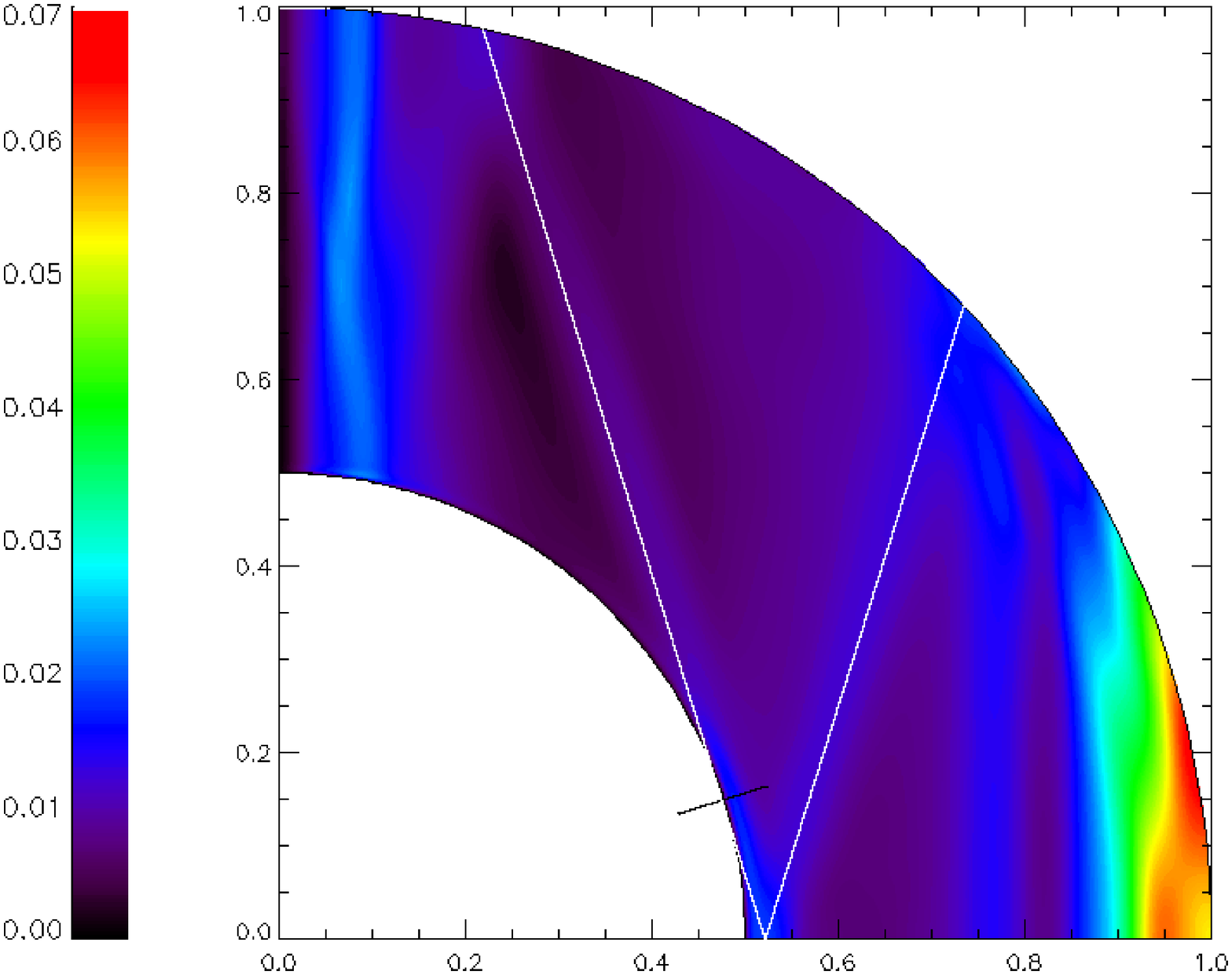}}
    \end{tabular}
  \caption{Azimuthally-averaged zonal velocity for $\omega/\Omega=1.1$ (top) and $\omega/\Omega=0.6$ (bottom). The amplitude of the forcing is $A=10^{-2}$ and $E=10^{-5}$. The boundary condition on the inner core is stress-free on the left and no-slip on the right. The poloidal component of the flow is very similar in both cases.\label{fig:compbc}}
\end{figure}

We show the azimuthal velocity in a meridional plane for $\omega/\Omega=1.1$ and $0.6$ and both boundary conditions in Fig.~\ref{fig:compbc}.
The time is the same in both cases, $\Omega t = 1225$.
The zonal flow is initially generated locally where the waves reflect from the outer boundary and then spreads in cylindrical radius due to viscosity.
In the no-slip case, the zonal flow only exists outside the cylinder tangent to the inner core whereas it continues to spread in the stress-free case.
Note however that for $\omega/\Omega=0.6$, a weak zonal flow is generated within the tangent cylinder in both cases.
This shows that a no-slip inner core has only a weak effect on the dissipation mechanism and generation of zonal flows for the parameters considered in this paper.
\subsection{Limitation of the model\label{sec:limit}}

\begin{figure}
  \begin{center}
    \begin{tabular}{c}
      \resizebox{80mm}{!}{\includegraphics{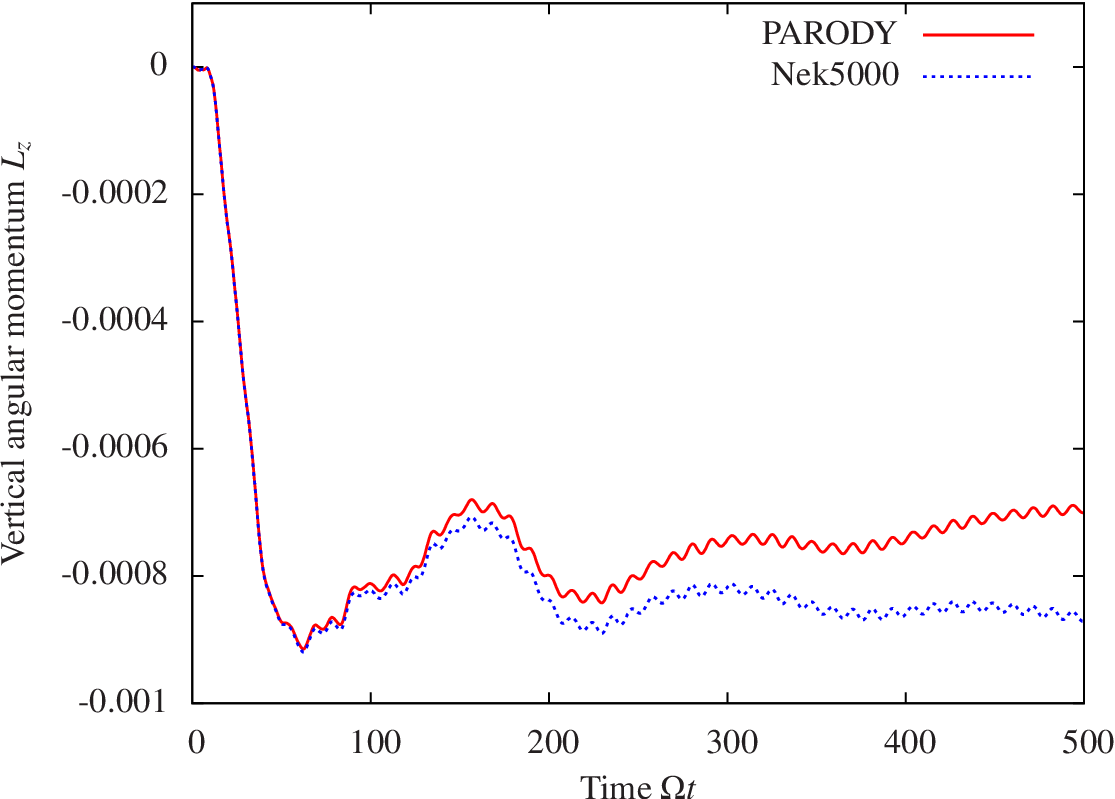}}
    \end{tabular}
    \caption{Vertical component of the angular momentum for $E=10^{-5}$, $A=10^{-2}$ and $\omega/\Omega=0.75$. Although the forcing frequency is positive, we observe a decrease in the spin frequency of the fluid. We show the results from both numerical methods.\label{fig:anom}}
  \end{center}
\end{figure}

When the Doppler-shifted frequency $\omega$ is positive, one would expect the fluid to spin up as the forcing injects angular momentum into the system.
This can be thought to represent the synchronisation of the spin of the body with the orbit of its companion when the companion is orbiting at a faster rate than the primary body is spinning, so that the tidal torque causes the angular momentum of the body to increase.
The opposite is supposed to happen for $\omega<0$, which corresponds to the case of a companion orbiting at a slower rate than the spin of the body, so that the tidal torque acts to spin down the body.
While this is true for most of the frequencies and amplitudes that we have investigated, some striking exceptions were observed.
This is the case when $\omega/\Omega=0.75$, for example.
In this case, the angular momentum should increase, whereas the opposite is observed in our simulations, leading to a desynchronisation of the fluid, which is unexpected.
\textcolor{black}{This anomalous evolution of the angular momentum is also observed for $\omega/\Omega=1.87$, see Fig.~\ref{fig:azimuth}, where the vertical angular momentum is mostly negative.}
We compare the time evolution of the vertical component of the angular momentum $L_z$ in Fig.~\ref{fig:anom}.
The results from both codes are shown for $E=10^{-5}$ and $A=3.68\times10^{-3}$.
Note that this unexpected result persists for other Ekman numbers and forcing amplitudes.
Although there is a slight disagreement between the codes at large times, the fact that the vertical component of the angular momentum is negative is a robust feature (the discrepancy between the codes appears to be caused by differences in the excitation of transients at $t=0$, as the forcing is instantaneously switched on).

This behaviour is observed for particular frequencies only.
It is not a result of numerical errors, since both codes exhibit the same behaviour, and the spatial and temporal resolution of these results has been carefully checked for convergence.
In addition, this does not appear to be only a transient phase in the simulations, and is observed to persist after several thousand rotation periods.
It is therefore a surprising property of the model for our nonlinear simulations at certain specific frequencies.

This behaviour is ultimately due to the fact that we consider an open system with a non-vanishing radial velocity at the outer boundary.
While we constrain the velocity of the fluid going in and out of the system to mimic the elliptical tidal deformation of the body, we do not constrain the ingoing (vertical) angular momentum flux.
The angular momentum injected in the system is related to Reynolds stresses $\left<u_ru_{\phi}\right>_{\phi}$ at the outer surface as described in Section \ref{sec:angmom}.
While the radial velocity is imposed at $r=r_e$, the azimuthal component is not constrained, since we adopt stress-free boundary conditions on a sphere instead of an ellipsoid.
This seems to be insufficient to realistically constrain the angular momentum evolution in some cases.
In addition, while we clearly observe a zonal flow driven by the wave reflection on the outer boundary, it is not clear to what extent this conclusion depends on our choice of boundary conditions.
Recent experimental results seem to be qualitatively consistent with our conclusions \citep{sauret,sauret2013b}, but the mechanism responsible for the zonal flows ought to be clarified in a more realistic model, either using the realistic ellipsoidal geometry, or introducing a body force in order to avoid the unrealistic angular momentum source term at the outer boundary.

These examples clearly illustrate the limitations of the current model.
Nevertheless, the qualitative behaviour of the nonlinear system described by our simulations is very likely to be robust, as the similitude between our results and the ones reported by \cite{tilgner2007} and \cite{sauret,sauret2013b} suggests.
This includes the generation of zonal flows, complicated time evolution of the dissipation rates with time and secondary shear instability, which are expected to all persist in a more realistic model.
 
%
\section{Discussion}\label{sec:conclusion}

In this paper, we have investigated numerically, for the first time, the nonlinear behaviour of tidally forced inertial waves in a spherical shell, as an initial value problem.
The forcing corresponds to an imposed radial velocity at the outer boundary, mimicking the radial displacement the fluid would experience in response to a tidal gravitational potential.
We have focused here on the dominant $l=m=2$ spherical harmonic, which is usually the dominant one for astrophysical applications, such as tidal synchronisation. This intentionally simplified model is designed to be an idealised representation of the convective regions of a giant planet, the fluid regions of a lower mass planet such as a Neptune mass planet, or to the convective envelope of a solar-type star.
This allows us to perform a detailed study of the nonlinear evolution of tidally forced inertial waves.

First, the purely linear regime is studied, and we recover previous results obtained using a direct solution of the steady-state response \citep{ogilvie2009}.
In the nonlinear regime, the forcing injects or extracts angular momentum (depending on the sign of the Doppler-shifted frequency $\omega$) in a non-uniform manner, leading to the generation of significant differential rotation in the interior of the body.
As the fluid spins up (or down), the fluid experiences highly time-dependent dissipation rates, which are a complicated function of the forcing frequency.
The amplitude of these zonal flows scales as $A^2E^{-\gamma}$ where $\gamma$ is some positive number depending on the frequency of the forcing.
This suggests possible shear instabilities in the low Ekman number or large amplitude regime that may be expected in the interiors of short-period extrasolar planets.
A hydrodynamical shear instability was indeed observed in some of our simulations, which is similar to an instability recently observed in laboratory experiments \citep{sauret,sauret2013b}.
Note however that we cannot at this stage perform a quantitative comparison between experiments and numerical simulations as there are significant differences between the two, the most important being the different type of boundary conditions used (stress-free plus inflow on a sphere in our simulations versus no-slip on a deformed sphere in experiments).

We have so far considered the regime of moderate Ekman numbers, which is the regime that is possible to access numerically, as well as relatively low amplitudes, to try to understand the dominant nonlinearities.
It is possible that this restriction constrains the nonlinearities to promote the importance of zonal flows.
This restriction might also eliminate the possibility of the internal wave beams becoming unstable to small-scale parametric instabilities.
At larger amplitudes, and more importantly, at lower Ekman numbers, parametric subharmonic instabilities, as recently observed by \cite{bordes2012} in the laboratory, or nonlinearities at the critical latitude itself \citep{GoodmanLackner2009}, might become more important, and might wash out some of the frequency dependence of the dissipation rate.
These possibilities will be explored in future studies. Nevertheless, the ultimate nonlinear outcome of these instabilities is likely to also result in the formation of zonal flows such as those that we have observed.

Due to the complexity of this problem and the limitations of our intentionally simplified model, we cannot currently compare our results directly to astrophysical observations.
However, it is worth mentioning those aspects that are likely to be relevant for the astrophysical problem.
When nonlinearities are considered, the flow can depart significantly from what would be predicted from linear theory, primarily due to the generation of differential rotation in the initially uniformly rotating body, in the form of zonal flows.
These zonal flows can play a role in modifying the subsequent dissipation rate, which is what is important in relating our work to astrophysical observations.
This is one example to illustrate that the simple picture assuming the body approaches synchronism as a uniformly rotating body is likely to be incorrect.
This is also true when tidal forcing is able to excite internal gravity waves \citep{GN1989,BarkerOgilvie2010}.
In all cases, the inclusion of nonlinearities and time evolution of the angular momentum leads to complicated time-dependence of the flow and the dissipation rate.
There are large differences between the dissipation rate as the amplitude and frequency of the tidal forcing is varied, as well as the Ekman number of the body.
There does not appear to be a general trend for the normalised dissipation rate: sometimes nonlinear terms increase the dissipation rate over the linear prediction, and they sometimes decrease it. 

Tidal forcing is likely to lead to complicated (cylindrical) differential rotation in the interior of the body.
Such differential rotation could be acted on by hydrodynamical or magnetohydrodynamical shear instabilities, or suppressed by magnetic stresses and convection.
Those secondary processes could play an important role by contributing to tidal dissipation in fluid bodies.
Future work will include studying more realistic mechanisms of tidal forcing, which do not exhibit the limitations of the current model discussed in Section~\ref{sec:limit}, in order to more accurately capture the tidal forcing of inertial waves in fluid stars and planets. In addition, topics of further study include studying the influence of stratification and convection, the latter of which could play a role in dissipating inertial waves, and the internal structure of the body and the imperfect rigidity of the core.

{\bf Acknowledgements} This work has been financially supported by STFC.
CPU time was provided by the UKMHD supercomputing facility located in Warwick.
BF thanks the Cambridge Newton Trust for financial support.
CB acknowledges support from a Herchel Smith Postdoctoral Fellowship of the University of Cambridge.
%
%
\bibliographystyle{mn2e}
\bibliography{biblio}
\label{lastpage}

\end{document}